%% file: paper_TOMM.tex
\documentclass[acmsmall]{acmart}
\AtBeginDocument{%
  \providecommand\BibTeX{{%
    \normalfont B\kern-0.5em{\scshape i\kern-0.25em b}\kern-0.8em\TeX}}}

\usepackage{graphicx}
\usepackage{algorithmicx}
\usepackage{verbatim}
\usepackage{subcaption}
\usepackage{caption}
\usepackage{hyperref}
\usepackage{lipsum,amsmath,multicol}
\usepackage{color,soul}
\usepackage{qtree}
\usepackage{xcolor}
\usepackage{enumitem}
\usepackage{algorithm}
\usepackage{algpseudocode}
\usepackage{pifont}
\usepackage{url}
\usepackage[usestackEOL]{stackengine}
\usepackage{fancyvrb}

\setcopyright{none}
\renewcommand\footnotetextcopyrightpermission[1]{} 
\begin{document}

\title{Neural Network Assisted Depth Map Packing for Compression Using Standard Hardware Video Codecs}
\author{Matti Siekkinen}
\affiliation{%
  \institution{Aalto University, Department of Computer Science}
  \country{Finland}
}
\affiliation{%
  \institution{University of Helsinki, Department of Computer Science}
  \country{Finland}
}
\email{matti.siekkinen@aalto.fi}

\author{Teemu Kämäräinen}
\affiliation{%
  \institution{University of Helsinki, Department of Computer Science}
  \country{Finland}
}
\email{teemu.kamarainen@helsinki.fi}

\renewcommand{\shortauthors}{Siekkinen and Kämäräinen}

\begin{abstract}
  Depth maps are needed by various graphics rendering and processing operations. Depth map streaming is often necessary when such operations are performed in a distributed system and it requires in most cases fast performing compression, which is why video codecs are often used. Hardware implementations of standard video codecs enable relatively high resolution and framerate combinations, even on resource constrained devices, but unfortunately those implementations do not currently support RGB+depth extensions. However, they can be used for depth compression by first packing the depth maps into RGB or YUV frames. We investigate depth map compression using a combination of depth map packing followed by encoding with a standard video codec. We show that the precision at which depth maps are packed has a large and nontrivial impact on the resulting error caused by the combination of the packing scheme and lossy compression when bitrate is constrained. Consequently, we propose a variable precision packing scheme assisted by a neural network model that predicts the optimal precision for each depth map given a bitrate constraint. We demonstrate that the model yields near optimal predictions and that it can be integrated into a game engine with very low overhead using modern hardware.
\end{abstract}



\maketitle

\input{01_intro}

\input{02_rw}

\input{03_depth_packing}

\input{04_neural_packing}

\input{06_performance}

\input{08_conclusion}


\bibliographystyle{ACM-Reference-Format}
\bibliography{biblio}

\end{document}

%% file: 01_intro.tex
\section{Introduction}
\label{sec:intro}

Depth is an important metric in graphics rendering and processing. A depth map is a representation of the distance of pixels to the camera that recorded the image. Depth maps are necessary, for example, in rendering to handle occlusions properly and in computer vision to perform 3d object detection. 

With the introduction of various distributed architectures for rendering and graphics processing, it has become necessary to to transmit depth information across a network efficiently. Depth image based rendering (DIBR) uses a color image supplemented with depth information in order to reproject that image to a new viewpoint. DIBR enables immersive video viewing with 6 degrees of freedom (6DoF) in which multiple video streams from different viewpoints can be reprojected to create a completely new view from perspective not directly provided by any of the individual cameras~\cite{boyce21immersive}. In remote rendered augmented (AR) and virtual reality (VR) where graphics are rendered by remote server and streamed as video to a thin client device, DIBR can be used to compensate for latency with 6DoF motion by streaming also depth and reprojecting incoming video frames to the current viewpoint of the client device~\cite{Shi2015}. A split rendering system presented in \cite{stengel21mmsys} streams remotely computed global illumination as compressed light probe data including depth enabling dynamic high-quality lighting on a resource constrained device, such as standalone virtual reality (VR) headset.

Similar to color videos, transmitting depth information requires fast performing compression, which is why video codecs are often used. Both, extensions for standard codecs as well as custom solutions for depth compression have been developed. However, currently none of these solutions are enabled with hardware accelerated encoding and decoding. We focus solely on solutions that are feasible for real-time encoding of rendered graphics using currently available (cloud) server hardware and for real-time decoding using currently available mobile devices, such as phones and standalone VR/AR devices. For this reason, using hardware implementations of standard codecs is the only feasible way currently to stream depth with high enough resolutions and frame rates. This in turn requires packing depth maps into YUV or RGB video frames prior to video compression because the RGB-D extensions for standard codecs are not supported by current hardware implementations. The dedicated piece of hardware for video coding also makes sure that the CPU and GPU of the device are not burdened with depth map encoding or decoding, as in a typical scenario they are required for other operations, such as graphics rendering at server side and warping the incoming depth images at client side.

In this paper, we investigate different ways of packing depth maps extracted from computer generated game scenes into color textures so that the resulting accuracy and bitrate tradeoff is as good as possible when the depth maps are compressed and decompressed using a standard video codec. We have used H.264 in this work but the methods can be applied to other video codecs with available hardware encoder and decoder implementations, such as H.265, as well. Specifically, we examine the combined impact of depth packing precision, i.e. how many bits are used to represent depth, and lossy video encoding on the resulting depth error when bitrate is constrained. Our results show that there is an optimal precision to use in depth map packing that depends on the target bitrate. In other words, while naive 8-bit depth packing (e.g., grayscale image encoding) provides too low precision considering that game engines can typically provide up to 24-bit depth when rendering to a texture, it is also generally not useful to pack depth using maximal number of bits (24) because the benefit of inter-frame compression diminishes with increasing precision.

The results also reveal that the optimal number of bits to use in depth map packing for video encoding is not constant but depends on the game scene and even varies between frames from the same scene. To cope with this, we adapt a neural network model to predict the optimal depth precision to use given a depth map and bitrate constraint. Evaluation results reveal that the model outperforms manually extracted baseline and produces near optimal predictions. The model generalizes for different trajectories of the same scene and the results suggest that it can also generalize across different scenes if provided with sufficiently diverse training data, alleviating the need to fine tune the model for each different scene/game. We benchmark the performance of the model in different scenarios and demonstrate that a prediction can be inferred in under a millisecond with a reasonable resolution using modern hardware and optimized inference engine. We also show that integrating the model into a game engine introduces only 0.5-2 ms of overhead depending on the hardware.

%% file: 02_rw.tex
\section{Related work}
\label{sec:rw}

Many different solutions have been proposed to compress depth maps over the years. Concerning standards, a 3D extension has been defined for the HEVC standard introducing several techniques to improve coding of depth maps~\cite{tech16hevc3d}. A number of improvements to its depth coding tools have also been proposed in recent years (e.g., \cite{lin2021visual,saldanha2019fast,shen2018efficient}). However, there is no support yet for hardware-accelerated coding using this standard, even though work on designing such efficient hardware has been conducted (e.g., \cite{sanchez20hevc3d,amish2019efficient}). Therefore, these solutions are unfeasible currently when high resolution and frame rate depth map streaming is required, which is the case with immersive applications, for instance. This is especially the case with mobile phones and standalone VR/AR headsets that have limited power and computational resources. 

The MPEG Immersive Video (MIV) specification is codec agnostic but the common test conditions of the working group used 10-bit depth with the HEVC Main 10 profile (i.e., grayscale images)~\cite{boyce21immersive}. In this way, hardware encoding and decoding are enabled with most modern graphics cards and embedded devices but the depth precision is currently limited to 10 bits unless a specific depth packing method is applied. An alternative approach to depth map encoding and streaming within MIV is to estimate depth from the received multiple different streamed views and even a compromise between the two is possible as shown in~\cite{garus2021immersive}. However, these solutions require multiple incoming video streams, and the depth estimation process is computationally much too demanding to be performed in real time for resource constrained mobile devices.

The light probe streaming solution presented in ~\cite{stengel21mmsys} packs depth information directly into YUV channels according to a predefined bit arrangement which is then streamed using lossless HEVC encoding, hence providing high performance also on mobile devices. While lossless encoding maintains depth maps intact, it is very inconvenient for all scenarios where bandwidth is limited because the resulting bitrate cannot be controlled. In this paper, we specifically examine the behaviour with lossy compression where the encoder rate control is given a specific bitrate target. This behaviour is typical for real-time streaming protocols, such as WebRTC, that need to react to varying available bandwidth in a frame-by-frame manner.

Another lossless depth encoding scheme is presented in \cite{wilson2017fast}. It is shown to perform fast compared to regular image compression alternatives on a PC and is a good candidate for such deployments. However, in addition to rate control being impossible, as a custom method it unfortunately does not enjoy the benefit of dedicated hardware acceleration as standard codec-based methods do, and it is therefore unlikely to be a feasible solution for high resolution and frame rate depth streaming to/from mobile, standalone devices.

In the telepresence systems presented in \cite{gunkel21mmsys,ekong2016teacher}, depth images are encoded as 8-bit grayscale images while in \cite{liu2015hybrid} the precision is set to 12 bits. All three solutions use the H.264 standard codec. The three proposed solutions use lossy, lossless, and hybrid lossless-lossy compression, respectively. The hybrid scheme compresses the most significant bits of depth images in lossless and the rest in lossy manner. In contrast to these works, we do not preset the depth precision but instead search for the optimal precision when applying lossy compression and given a bitrate target.

A custom depth image and video codec is proposed in \cite{koniaris18gpudepth}. The solution targets fast decoding by running the decoder entirely on GPU. However, the fixed 4:1
compression ratio provided by the solution is modest and does not allow dynamic rate control. In addition, as it is only benchmarked with many parallel depth streams on a PC with relatively powerful GPU, it is unclear how it would perform on resource constrained devices decoding a single depth stream compared to hardware accelerated standard codec.

Most closely related work to ours is presented in~\cite{pece2011adapting} which introduces a scheme to pack 16-bit depth into YUV textures for compression with standard codec. The method aims to reduce quantization and chroma subsampling effects as much as possible but provides a knob that can be turned to adjust the bitrate accuracy tradeoff. Unofrtunately, the paper does not explore the impact of adjusting that parameter at all. In this work, we use their depth packing scheme in some of our experiments in Section~\ref{sec:depth_packing} and examine the bitrate-accuracy tradeoff when adjusting the packing scheme parameter. The neural network assisted parameter optimization approach we propose can be applied to their packing scheme too.

%% file: 03_depth_packing.tex
\section{Depth map packing for standard video encoding}
\label{sec:depth_packing}

In this section, we present two different packing schemes and investigate what kind of bitrate vs. depth accuracy tradeoff they provide when used in conjunction with a standard H.264 video codec.

\subsection{Packing schemes}

A depth map consists of a value per pixel and the values can describe a linear or nonlinear distance of the pixel from the camera. They can also describe disparity in case of multiple cameras (e.g., stereo view). In order to pack a depth map into a color texture, the values need to be normalized. Game engines limit the depth range that cameras see by introducing so called near and far planes. Any object closer than the near plane or farther than the far plane will not be visible. The values stored in a depth buffer of a game engine renderer are typically nonlinear and normalized to $[0,1]$ range so that zero and one correspond to the depth of the near and far planes, respectively. Depth value normalized in this way is in practice stored into a 32-bit single-precision floating-point number (i.e., FP32) or an unsigned integer so that it can hold the depth at sufficient precision. Unity, for instance, provides up to 24-bit precision depth when rendering to a texture.

When packing a depth map for video compression using a standard codec, depth of each pixel is typically packed directly into the YUV channels so that the resulting texture is directly submitted to the video encoder without RGB to YUV color space conversion. We consider both YUV444 format, i.e., each channel is represented by 8 bits, and YUV420 format, i.e., the chroma channels UV are subsampled using 2x2 pixel blocks, which effectively reduces the number of bits per pixel into 12. Packing depth directly to YUV makes it easier to ensure that the most significant bits of the depth value are not affected by possible chroma subsampling by placing them to the Y channel.

\subsubsection{Variable bit packing (VBP)} VBP enables packing depth with different precision, i.e. different number of bits. When a full 24-bit precision is requested, the VBP scheme straightforwardly packs the bits into the YUV channels so that the most, the second most, and the least significant 8 bits are stored to Y, U, and V channels, respectively. In other cases, the depth float is first quantized according to the requested precision. For example, when packing using 12-bit precision, the first 8 bits are packed to the Y channel and the remaining 4 bits as the most significant bits of the U channel, leaving the rest of the bits of that channel as well as the whole V channel unused. In addition, the values packed to the U and V channels are inversed modulo two of the Y and U channels, respectively, effectively producing triangle waves with linearly increasing depth. The reason for this is to avoid discontinuities with continuously increasing and decreasing depth, which would be subject to more pronounced impact of quantization by the video encoder. The shader function code for the packing is shown in Algorithm \ref{algo:vbp}. The unpacking function (not shown for space constraints) simply performs the inverse operations.

\begin{algorithm}[t]
  \caption{Variable bit packing (VBP)}
\begin{Verbatim}[fontsize=\scriptsize]
inline float3 VariableBitPacking(float depth, uint bits) {
    if (depth == 1.0f) {
        return float3(1.0f, 1.0f, 1.0f);
    }
    float4 scale = float4(1.0f, 255.0f, 65025.0f, 16581375.0f);
    float3 ogb = float3(65025.0f, 255.0f, 1.0f) / 16581375.0f;
    float4 unit = float4(depth.xxxx);
    unit.gba -= floor(float3(unit.gba / ogb)) * ogb;
    float4 color = unit * scale;
    color = color - floor(color);
    color.rgb -= color.gba / 255.0f;

    //invert values to make triangle waves
    uint chromacbint = uint(round(float(color.g * 255.0f)));
    uint lumaint = uint(round(float(color.r * 255.0f)));
    if (chromacbint % 2 > 0) {
        color.b = 1 - color.b;
    }
    if (lumaint % 2 > 0) {
        color.g = 1 - color.g;
    }
    
    //reduce precision to specified number of bits
    float bits_to_use = min(8.0f, bits - 8.0f);
    if (bits_to_use > 0) {
        color.g = color.g * pow(2.0f, (bits_to_use - 8));
    }
    bits_to_use = min(8.0f, bits - 16.0f);
    if (bits_to_use > 0) {
        color.b = color.b * pow(2.0f, (bits_to_use - 8));
    }
    if (bits == 8) {
        color.g = 0;
    }
    if (bits <= 16) {
        color.b = 0;
    }
    return color.rgb;
}
\end{Verbatim}
\label{algo:vbp}
\end{algorithm}  

\subsubsection{Robust packing (RP)} RP refers to the packing scheme presented in~\cite{pece2011adapting}. It has been designed to be robust against compression artefacts due to quantization and chroma subsampling. The scheme packs 16-bit depth maps so that an 8-bit coarse depth is stored into Y channel and the chroma channels carry partially redundant information encoded as triangle waves with same period but different phase. The period of the triangle waves can be adjusted using a parameter $n_p$ and tuning it provides a way to allow 
the packed depth map to gradually become more and more subject to quantization.

\subsection{System setup}

In all the experiments, we used the Unity game engine supplemented with the WebRTC plugin from the Render Streaming package for real-time video streaming of the rendered graphics. We modified the plugin to submit textures in the chosen YUV format to the video encoder, to override the WebRTC congestion control feedback to the encoder rate control so that it obeyed our bitrate target, and to dump the encoded video to a file. Depth maps were created by injecting a custom pass to the HD Render Pipeline. The custom pass uses a shader to directly pack the camera depth buffer into a render texture that is subsequently passed to the video encoder plugin. The depth buffer contained inversed non-linear depth normalized to $[0,1]$ range according to the near and far plane, which were set to 0.1 and 100, respectively. This way of packing and encoding depth maps can be easily done in real-time at our target frame rates because the data does not leave GPU memory until it is read as a compressed video frame. 

In addition, two other forms of depth maps were gathered. Non-packed high precision depth maps were extracted as the ground truth data by using a compute buffer to store the depth buffer values, reading that buffer to CPU memory, and storing it as bit array to a file. For a subset of frames, packed but uncompressed depth maps were extracted as reference by saving the render textures as PNG image files before they are submitted to the video encoder. Afterwards, to evaluate the quality of the resulting compressed depth maps, video dumps were decoded and/or unpacked, and compared to the reference and ground truth data using FFmpeg, OpenCL, and Python scripts.

For the video encoding, we used hardware accelerated H.264 codec provided by the Nvidia Video Codec SDK 11.0 run by the Titan V graphics card. The encoder was configured to use P1 preset, CBR rate control, ultra-low latency tuning info, and other settings according to the recommendation by the Nvidia programming guide for "low-latency use cases like game-streaming, video conferencing etc."\cite{nvidia_codecguide}. The only exception to the recommendations were that we used one second long GOP (GOP length equal to the \texttt{frameRateNum} parameter in NVEncoder API with \texttt{frameRateDen}=1).

\subsection{Scenes and datasets}
\label{subsec:datasets}

\begin{figure}[t]
        \begin{subfigure}[b]{0.245\columnwidth}
            \centering
            \includegraphics[width=\columnwidth]{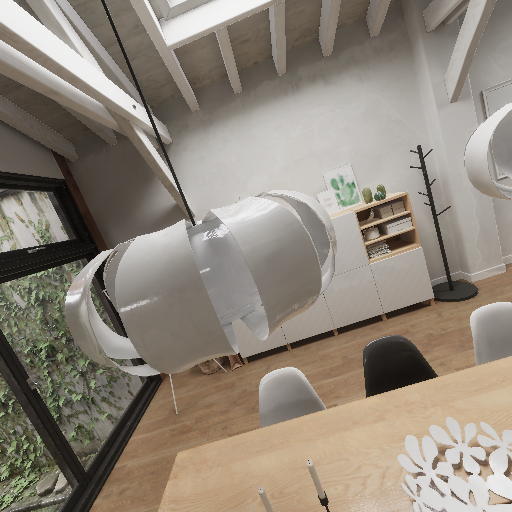}
            \caption{RGB capture}
            \label{fig:sample_rgb}
        \end{subfigure}
        \begin{subfigure}[b]{0.245\columnwidth}  
            \centering 
            \includegraphics[width=\columnwidth]{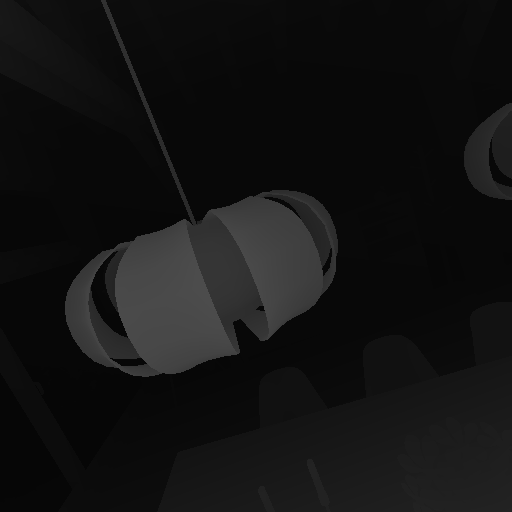}
            \caption{8-bit depth map} 
            \label{fig:sample_gs}
        \end{subfigure}
        \begin{subfigure}[b]{0.245\columnwidth}   
            \includegraphics[width=\columnwidth]{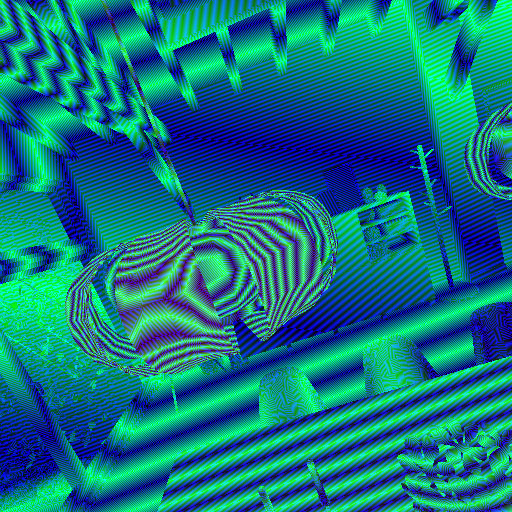}
            \caption{VBP w/ 24-bit precision}
            \label{fig:sample_VBP}
        \end{subfigure}
        \begin{subfigure}[b]{0.245\columnwidth}   
            \includegraphics[width=\columnwidth]{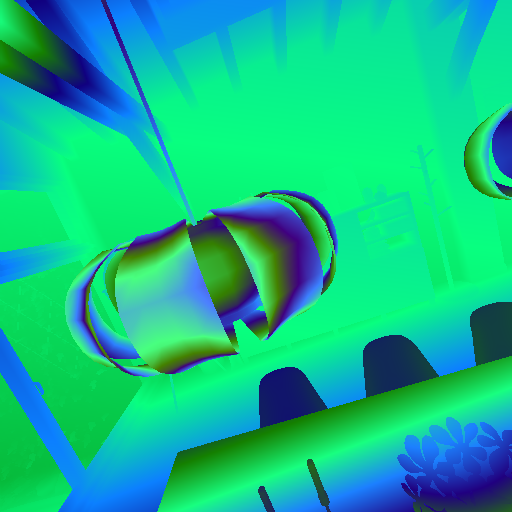}
            \caption{RP with $n_p=4096$} 
            \label{fig:sample_RP} 
        \end{subfigure}
        \caption{Sample frame from the Scandinavian House demo scene and corresponding depth map packed in three ways.} 
        \label{fig:samples} 
\end{figure}
    
To generate all the results presented in Section \ref{subsec:error}, we used a photorealistic architectural visualization scene from Oneiros~\cite{OneirosVR} called \textit{The Scandinavian House demo (AVP Vol.6)} available in the Unity Asset Store~\cite{archvizassetstore}. The scene is a fully navigable interior of a Scandinavian house that includes more than 200 objects and 4K textures. The scene was adapted for VR and the target framerate set to 90 fps. We recorded the camera trajectory (pose including rotation and position) while a user explored the different rooms in the virtual house for a total of 6 minutes using the Oculus Quest VR headset. The user was able to teleport to fixed positions located inside the house in different rooms and look around with 6DoF. Figure \ref{fig:samples} displays a sample frame from this scene and the corresponding depth map packed in different ways.

The recorded trajectory was then repeatedly replayed so that each time either the ground truth depth data, packed reference depth maps, or compressed depth maps were grabbed with different depth packing and video encoding configurations. Temporal anti-aliasing was turned off to avoid depth jittering and all animations were disabled to have identical depth maps each time for packing and encoding. The replay resulted in compressed and ground truth depth maps corresponding roughly to 26k rendered frames of which roughly 3200 were also extracted as packed but uncompressed reference frames. We name this dataset \textit{HOUSE-VR}.

We also generated another dataset using the same scene with a prerecorded trajectory created with the Camera Path Creator asset for Unity. The resulting camera dynamics are very different from the VR usage and are closer to first person shooter (FPS) gaming as there is no head motion involved. The resulting dataset comprises depth maps corresponding to about 11k rendered frames. We name this dataset \textit{HOUSE-FPS}.

In addition, we created datasets from two other scenes, namely the Fontainebleau demo scene~\cite{fontainebleau} and the Unity HD Render Pipeline template scene \cite{hdrp_template}. Both were used with an FPS camera controller and played for a few minutes by moving and looking around the scene resulting to depth maps corresponding to roughly 10k rendered frames. We name these two datasets \textit{FBLEAU} and \textit{HDRP}.

\subsection{Impact of bitrate on depth encoding error}
\label{subsec:error}

\begin{figure}[t]
        \begin{subfigure}[b]{0.495\columnwidth}
            \centering
            \includegraphics[width=\columnwidth]{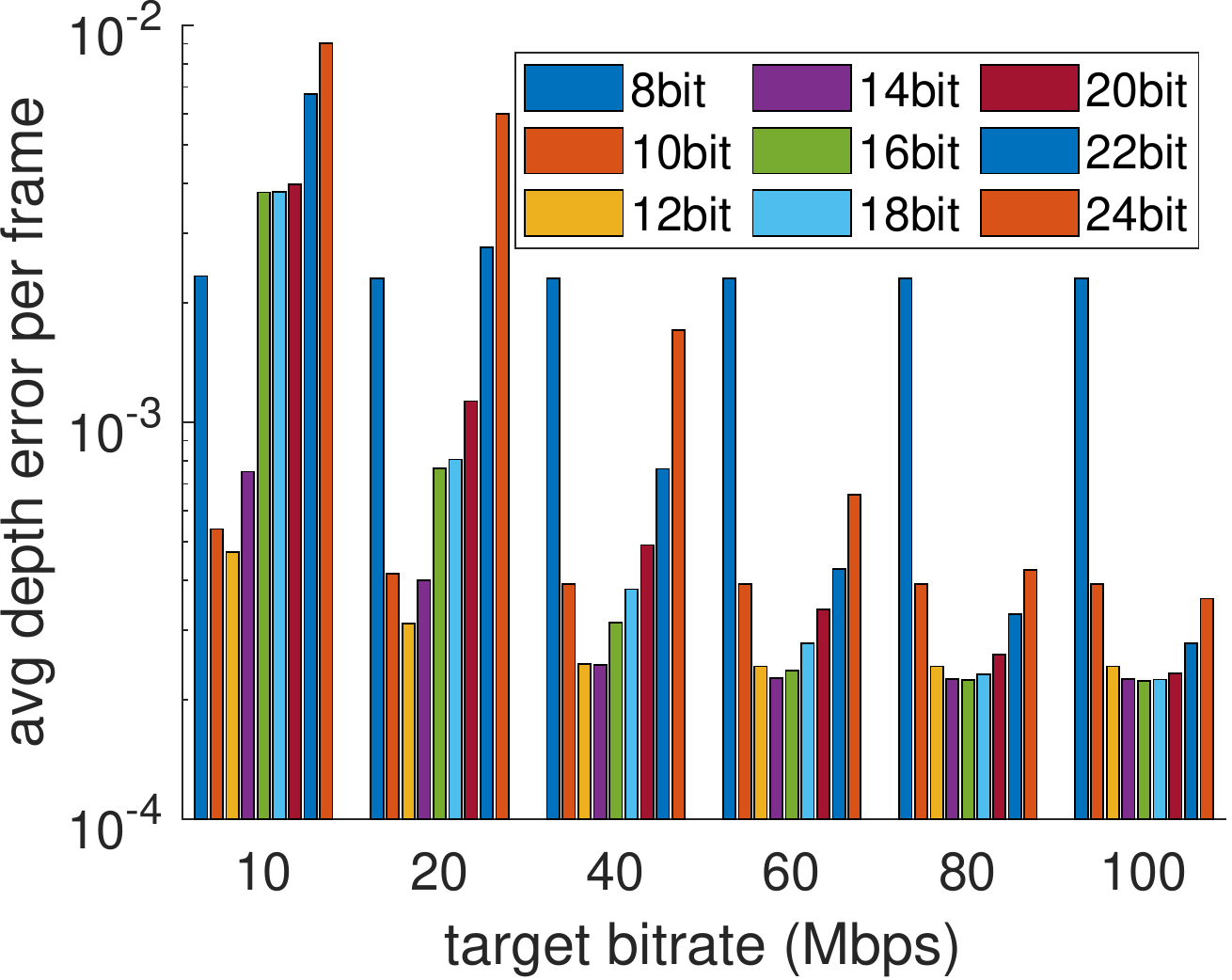}
            \caption{VBP with YUV420}
            \label{fig:mean_errors_vbp}
        \end{subfigure}
        \begin{subfigure}[b]{0.495\columnwidth}  
            \centering 
            \includegraphics[width=\columnwidth]{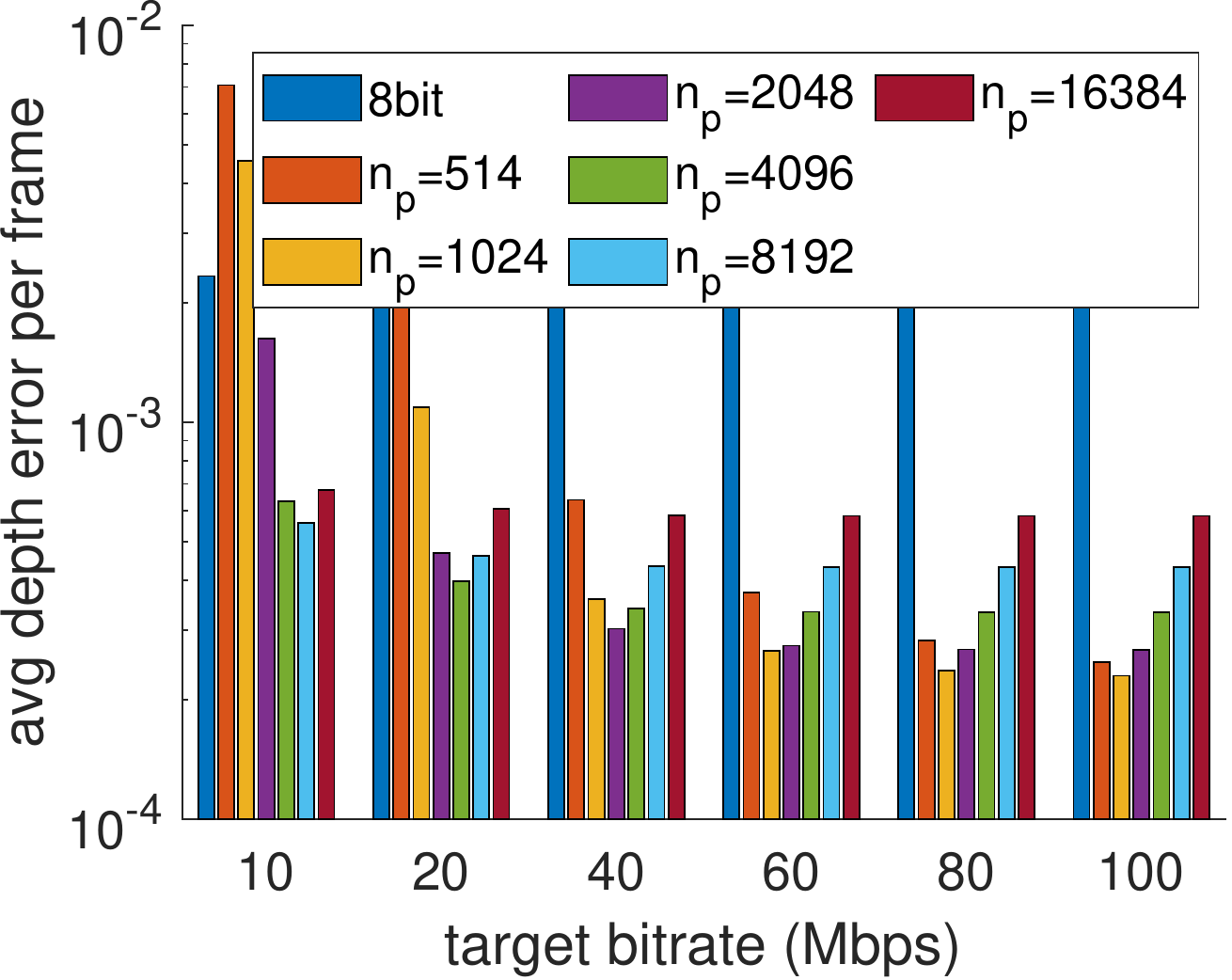}
            \caption{RP with YUV420} 
            \label{fig:mean_errors_rp}
        \end{subfigure}
        \vskip\baselineskip
        \begin{subfigure}[b]{0.495\columnwidth}   
            \includegraphics[width=\columnwidth]{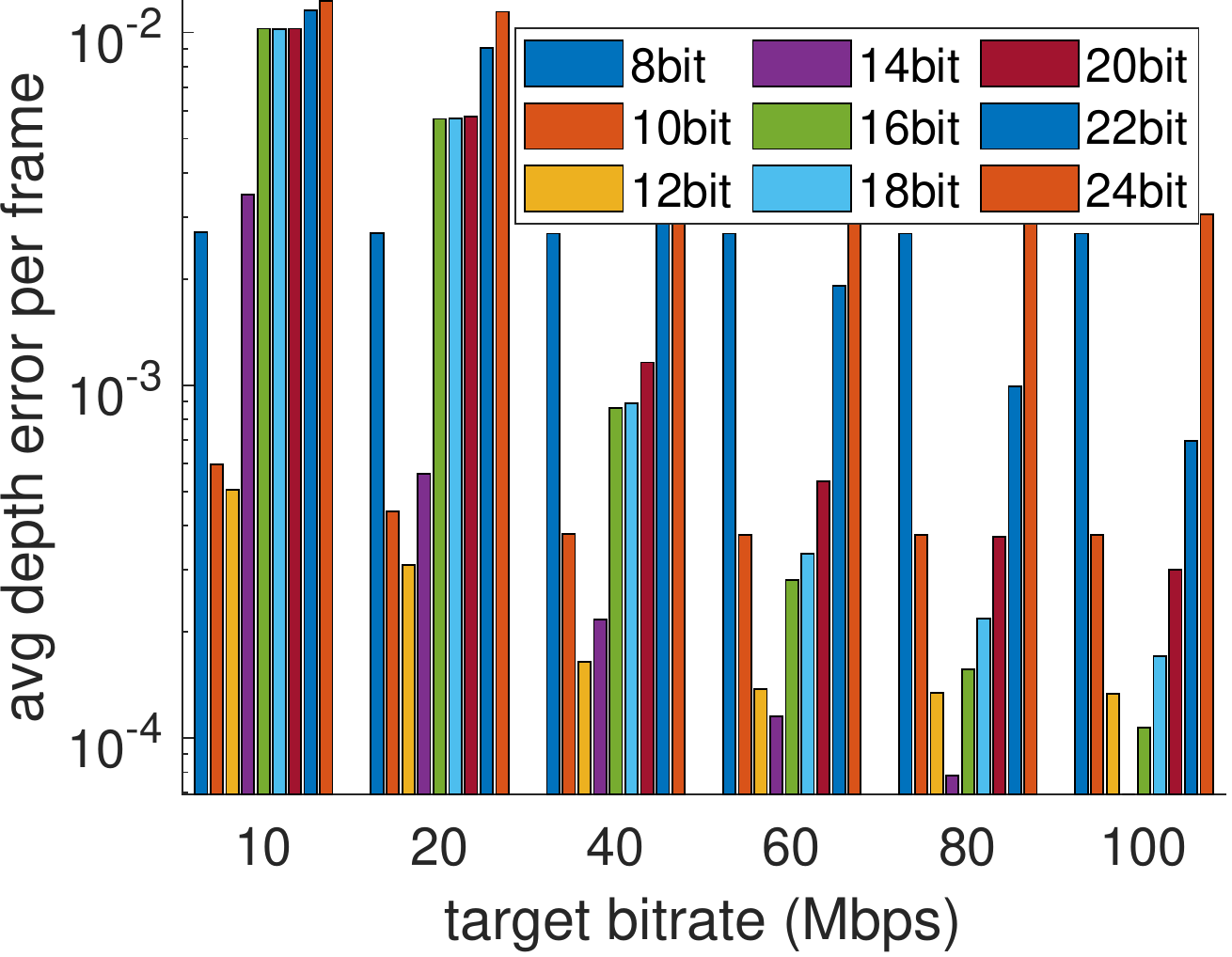}
            \caption{VBP with YUV444}
            \label{fig:mean_errors_vbp_444}
        \end{subfigure}
        \begin{subfigure}[b]{0.495\columnwidth}   
            \includegraphics[width=\columnwidth]{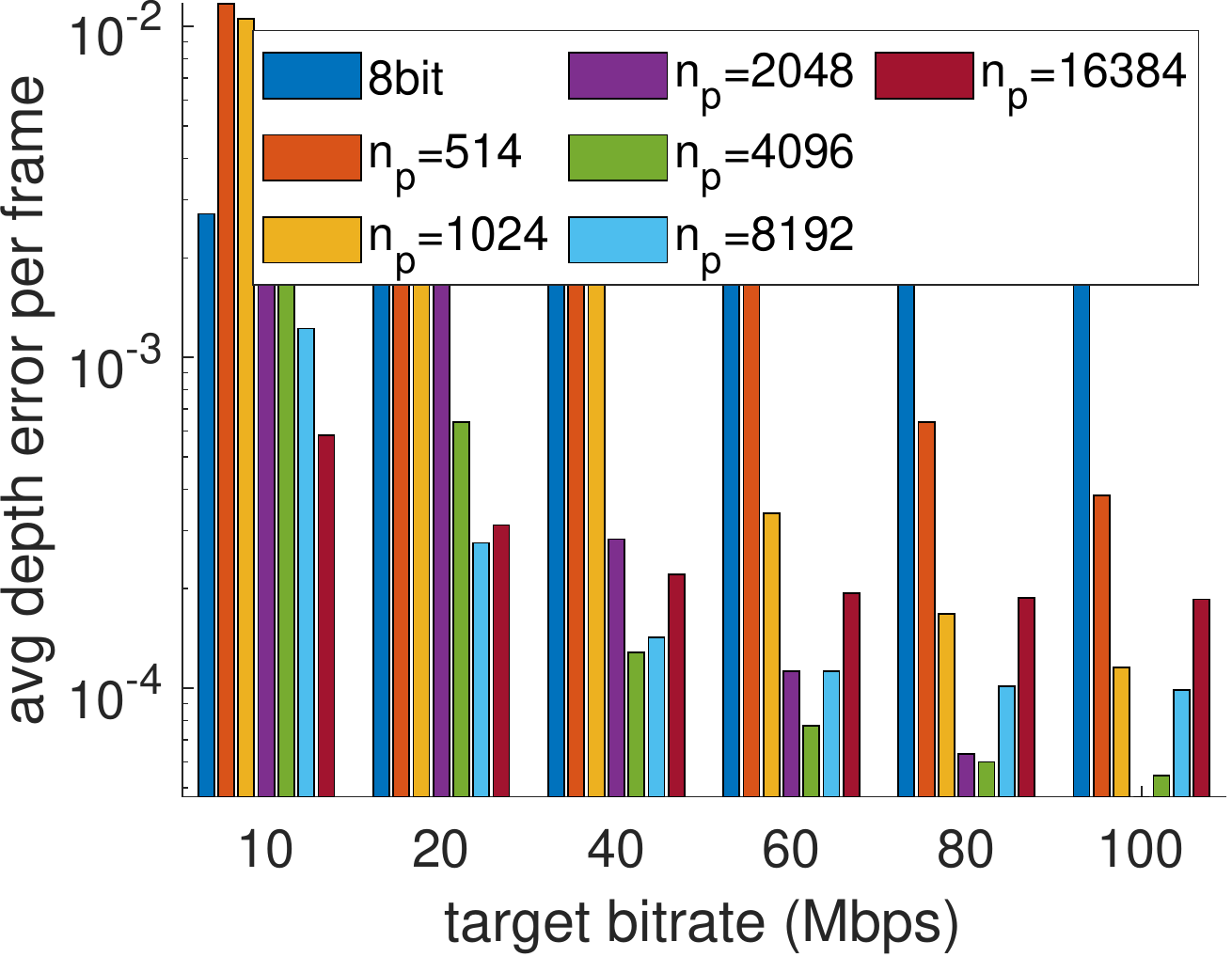}
            \caption{RP with YUV444} 
            \label{fig:mean_errors_rp_444} 
        \end{subfigure}
        \caption{Depth error with packing schemes and target bitrates.} 
        \label{fig:mean_errors} 
    \end{figure}


We now study what is the resulting error of a stream of depth maps that has been packed using one of the two schemes with given parametrization and encoded to video using specific target bitrate. 

Figure \ref{fig:mean_errors} shows the resulting mean depth map error averaged over all the frames for the different packing configurations and target bitrates. We show separately results with subsampled (YUV420) and full rate sampled (YUV444) chroma channels. 8-bit packing is included as well, and it corresponds to grayscale packing in which the depth value is replicated to all the three color channels. 

The most important observation is this: Increasing precision, i.e., the number of bits to use for packing with VBP, does not automatically result in reduced error after video compression and decompression. While increasing the precision reduces the error caused by the packing scheme itself, the error caused by quantization by the video encoder starts to increase after some point when the target bitrate remains constant. The reason for this is that it becomes increasingly difficult for the encoder to leverage inter-frame compression because small depth differences lead to relatively large differences in packed depth when using high precision. This means that there is a precision that provides the smallest error and it depends on the target bitrate. The RP scheme behaves similarly when we adjust the $n_p$ parameter, which means that there is an optimal value for that parameter, and it similarly depends on the target bitrate. Chroma subsampling is harmful with both packing schemes and full rate sampling improves the results especially with higher target bitrates.

\begin{figure}[t] 
\centering
  \begin{subfigure}[b]{0.495\linewidth}
     \includegraphics[width=1\linewidth]{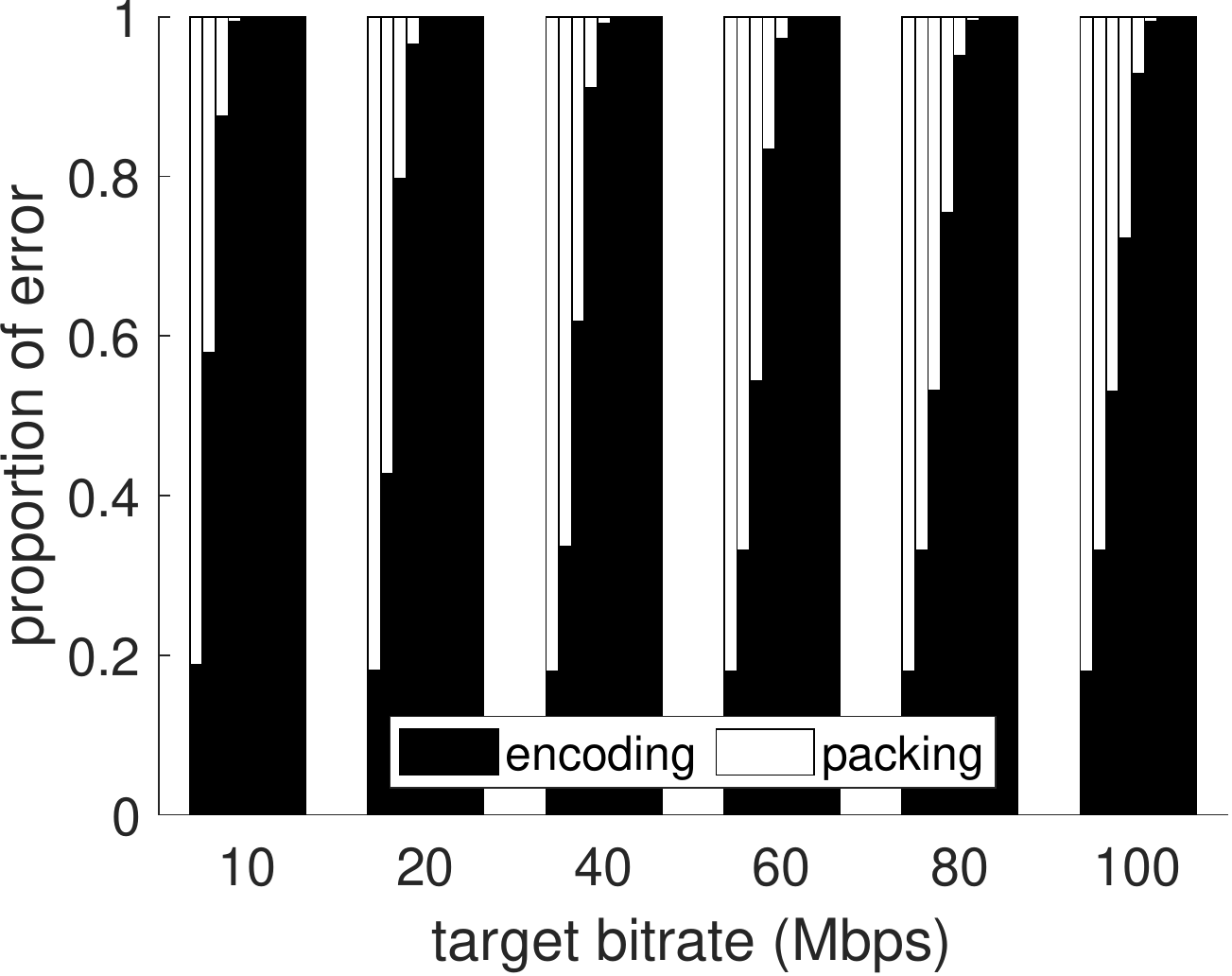}  
    \caption{VBP}
    \label{fig:error_proportion_vbp}
  \end{subfigure}
  \begin{subfigure}[b]{0.495\linewidth}
   \includegraphics[width=1\linewidth]{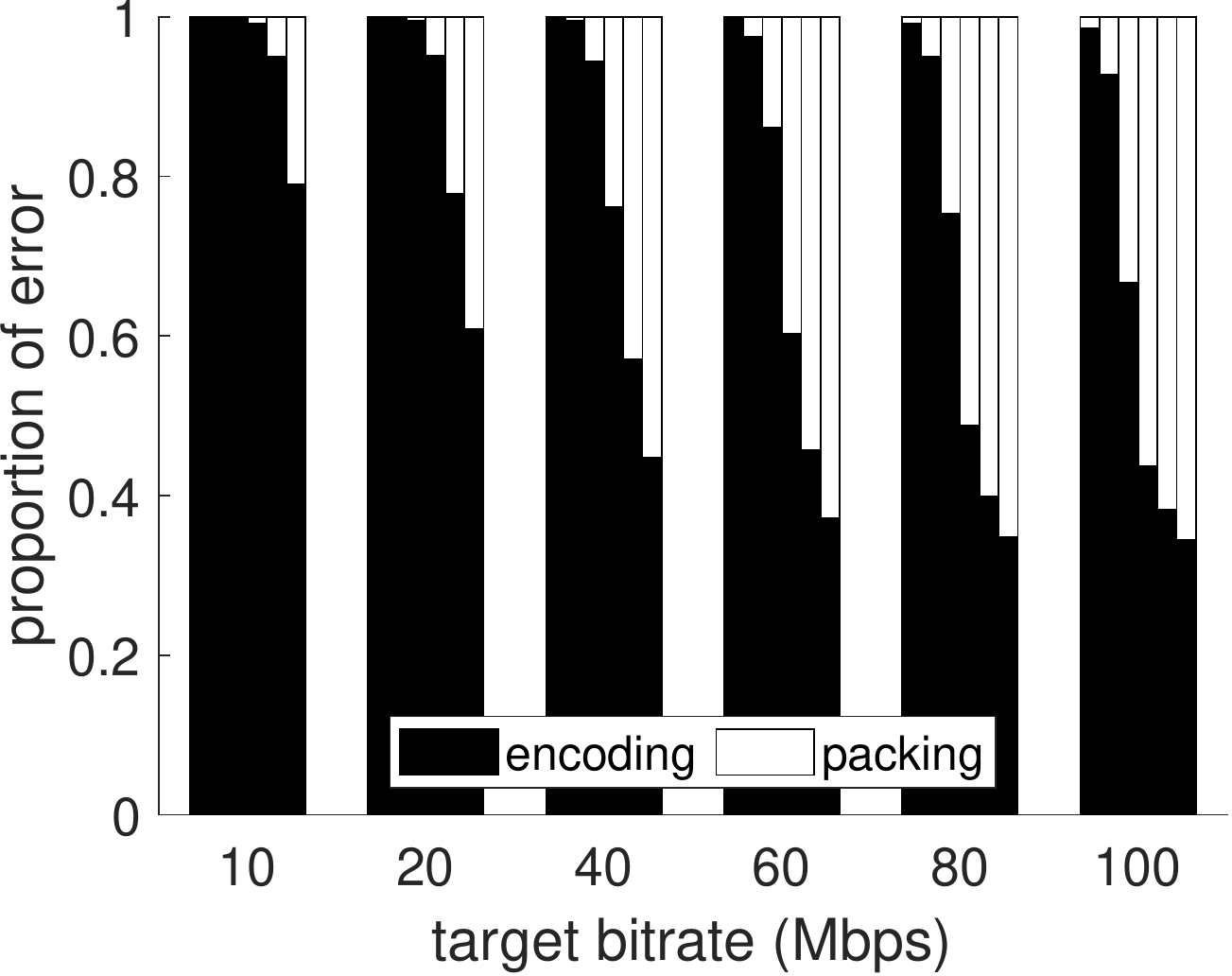}
    \caption{RP} 
    \label{fig:error_proportion_rp} 
    \end{subfigure}
  \caption{Error composition. Each bar group includes same ranges of depth bits for VBP and of $n_p$ for RP as in Fig. \ref{fig:mean_errors}.}
  \label{fig:error_proportion}  
\end{figure}

Figure \ref{fig:error_proportion} further illustrates what the resulting depth error is composed of in different scenarios (we only show YUV444 case as chroma subsampled is similar). When packing precision is increased, the proportion of accuracy lost by the packing scheme decreases and error by video encoding increases. The same applies to the Robust scheme when the $n_p$ parameter is decreased. The best configuration of a particular packing scheme therefore optimally balances these two sources of error.

\begin{figure}[t] 
\centering
  \begin{subfigure}[b]{0.495\linewidth}
     \includegraphics[width=1\linewidth]{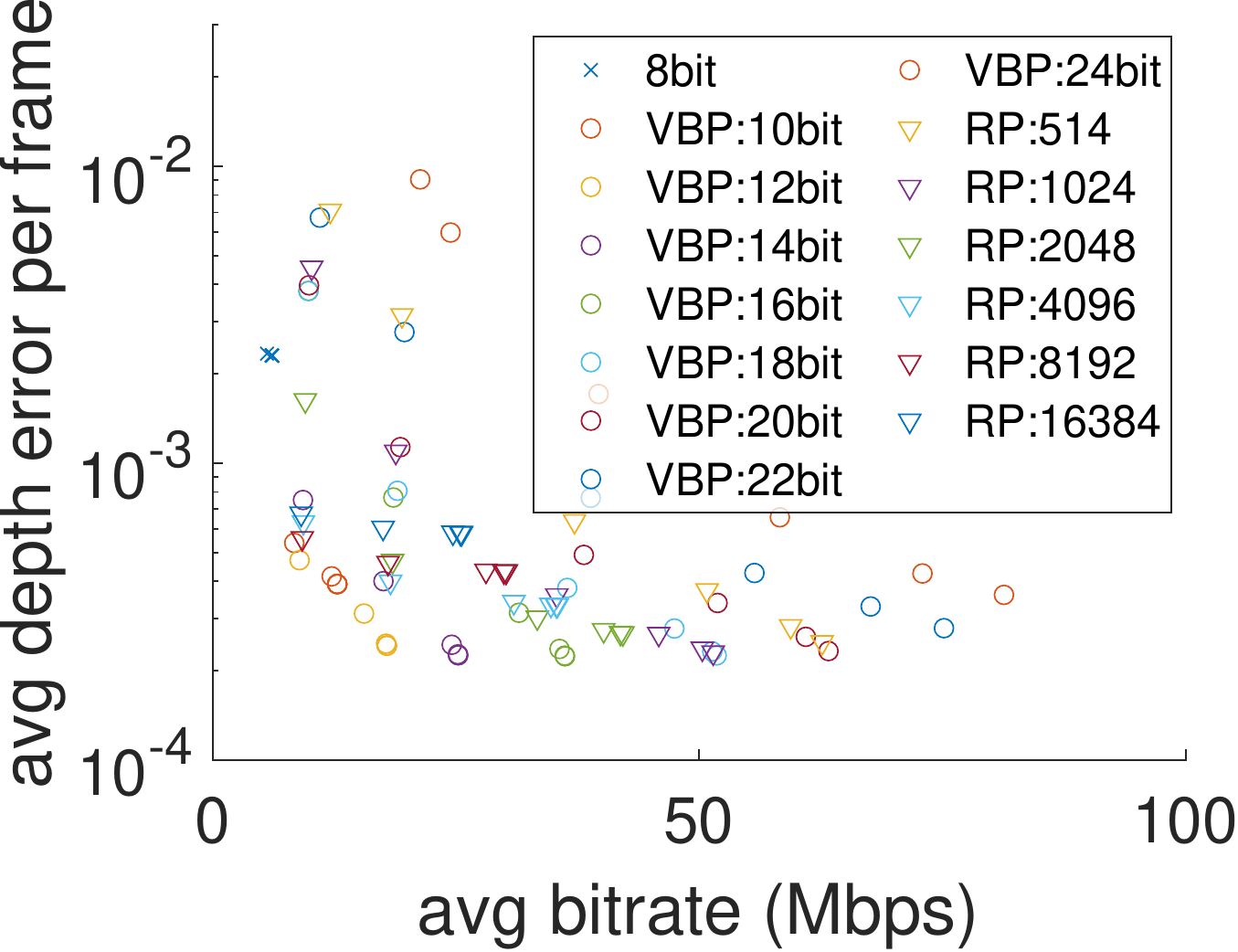}
    \caption{YUV420}
    \label{fig:error_vs_bitrate_420}
  \end{subfigure}
  \begin{subfigure}[b]{0.495\linewidth}
   \includegraphics[width=1\linewidth]{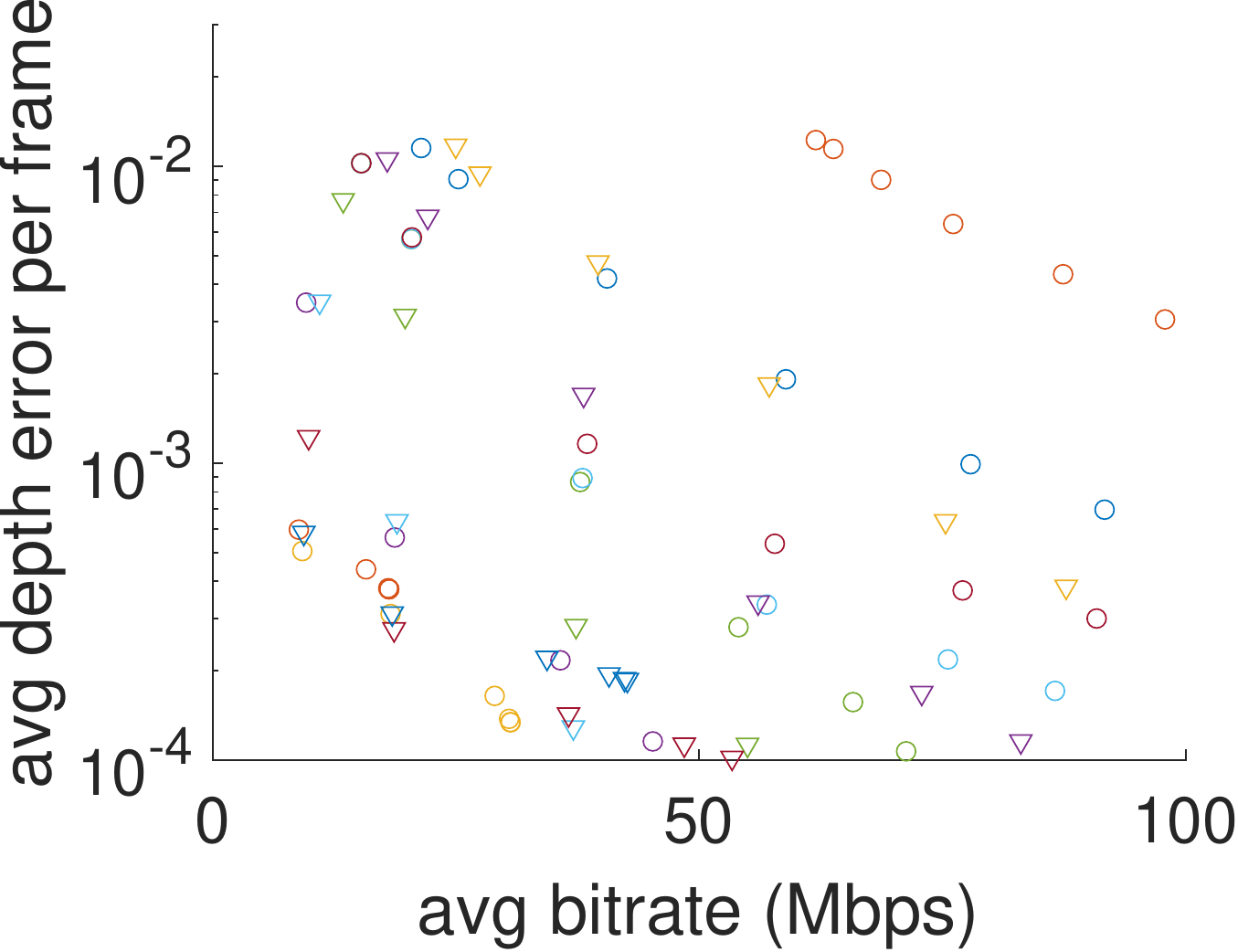}
    \caption{YUV444} 
    \label{fig:error_vs_bitrate_444} 
    \end{subfigure}
  \caption{Depth encoding error vs. average bitrate.}
  \label{fig:error_vs_bitrate}  
\end{figure}

We examine the tradeoff between depth encoding error and bitrate in Figure \ref{fig:error_vs_bitrate}, which plots the error against the resulting average bitrate (rate control may sometimes end up delivering bitrates that somewhat differ from the target). The lowest errors are achieved with the VBP when chroma channels are subsampled whereas RP works slightly better without chroma subsampling, but there is no clear difference between the two schemes.

\subsection{Depth map specific error}

\begin{figure}[t] 
\centering
  \begin{subfigure}[b]{0.495\linewidth}
     \includegraphics[width=1\linewidth]{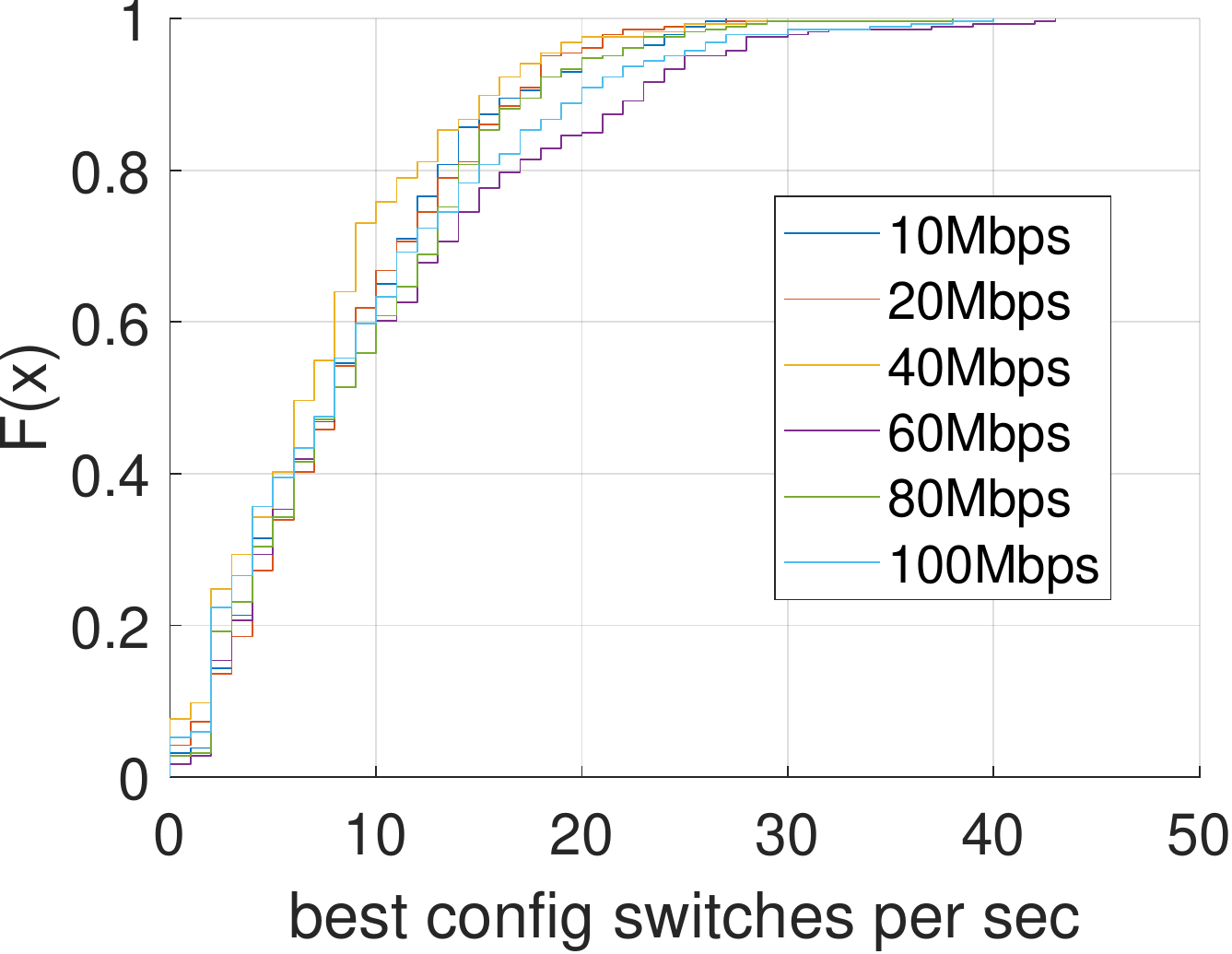}
    \caption{Best configuration stability.}
    \label{fig:best_config_switches}
  \end{subfigure}
  \begin{subfigure}[b]{0.495\linewidth}
   \includegraphics[width=1\linewidth]{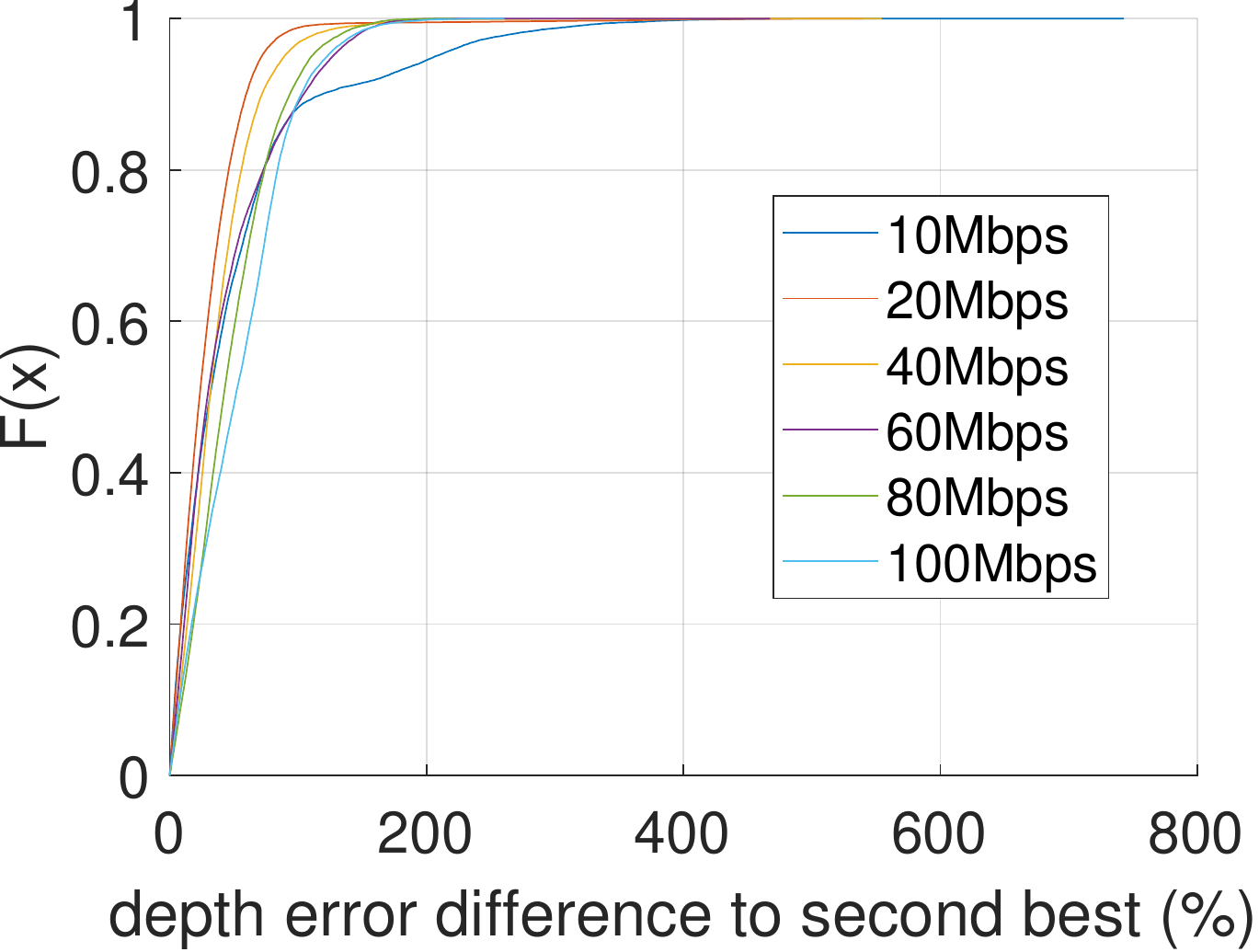}
    \caption{Second-best vs best.} 
    \label{fig:2nd_best_error} 
    \end{subfigure}
  \caption{Best depth packing configuration for given bitrate varies across depth maps.}
  \label{fig:best_config_plots}  
\end{figure}

So far, we have established that there is a target bitrate dependent configuration that minimizes the overall depth encoding error for a specific depth packing scheme when we compute the error over all the depth maps of the stream. We now look at how the error varies between individual depth maps.

We computed the depth packing configuration (both VBP and RP considered) that yields the smallest depth encoding error for each depth map with a given target bitrate. Figure \ref{fig:best_config_switches} shows how frequently it changes within the depth map stream. Figure \ref{fig:2nd_best_error} shows that the difference in the encoding error between the best and second-best configurations can be significant. The takeaway from these two graphs is that the optimal configuration is not the same for each depth map even if target bitrate stays constant and that there may be a substantial difference in encoding error between the best and second-best configuration. This result suggests that a depth map specific packing scheme would be useful.






%% file: 04_neural_packing.tex
\section{Neural network assisted packing}
\label{sec:model}

Inspired by the findings that there is an optimal packing configuration for given target bitrate and, furthermore, that this configuration is depth map specific, we design a model to predict the best configuration prior to the actual packing. We adapt an existing neural network model to this task and train it to predict the depth encoding error with different configurations. We consider here only the VBP scheme and YUV444 encoding but the model could as well be trained to predict the optimal $n_p$ parameter for the RP scheme and YUV420 encoding.

\subsection{Model}

\begin{figure}[t] 
    \centering
    \includegraphics[width=0.6\columnwidth]{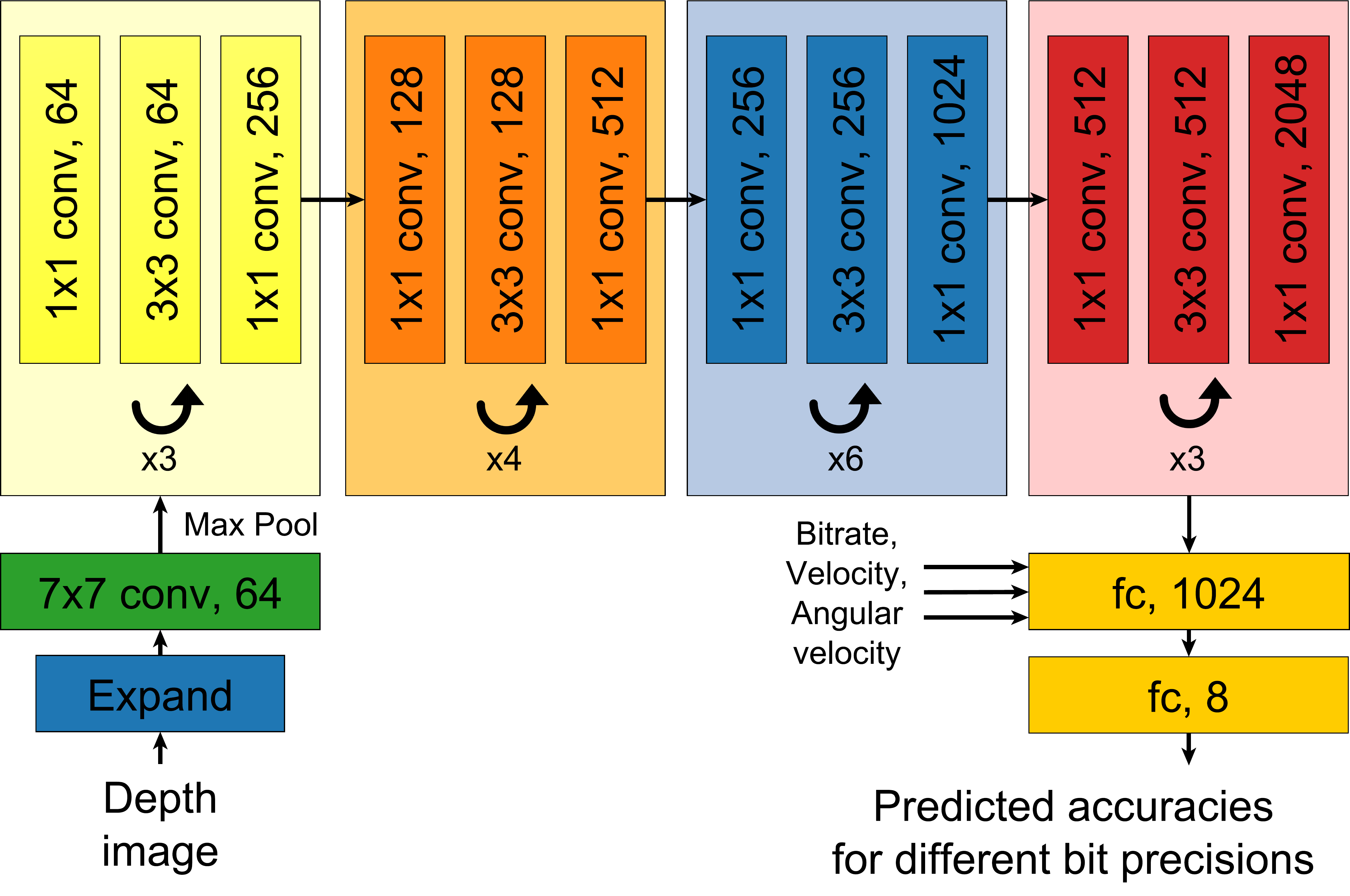}
    \caption{Modified ResNet-50 model for depth precision prediction. Inputs are depth image, bitrate, (camera) velocity and angular velocity. The model outputs the predicted accuracies for different bit precisions.}
    \label{fig:model}
\end{figure}

We repurposed the residual neural networks (ResNet) architecture for depth map encoding error prediction model. ResNet architecture was introduced by He et al.~\cite{he2016deep} and popularized the use of deeper architectures compared to previous work. This was made possible by using skip connections which reduce the information loss in deep networks. We found the ResNet-50 to be an optimal network depth for our task. The final model is illustrated in Figure~\ref{fig:model}.

The ResNet model was modified to perform multi-output regression instead of classification by using L1 loss function (i.e., depth error) and outputting predictions for each depth precision. We also added an expansion to the beginning of the architecture to expand a one channel depth image into three channels so that we could utilize pre-trained weights of networks trained with RGB input. The output of the model is the predicted encoding error for 8 different depth precision alternatives.

In order for the model to be able to predict the optimal precision with different target bitrates, we added a fully connected layer to the end of the network which also takes target bitrate of the video encoder as input. In addition, camera velocity and angular velocity (calculated using the previous and current depth map) as additional inputs to the layer. The reason for including velocities is to capture camera motion that affects the inter-frame compression of the video encoder. By training a model without the velocities as well, we confirmed that they provide useful information to the neural network and somewhat improve the prediction accuracy. 

\subsection{Training}

\begin{figure}[ht] 
\centering
  \begin{subfigure}[b]{0.4\linewidth}
     \includegraphics[width=1\linewidth]{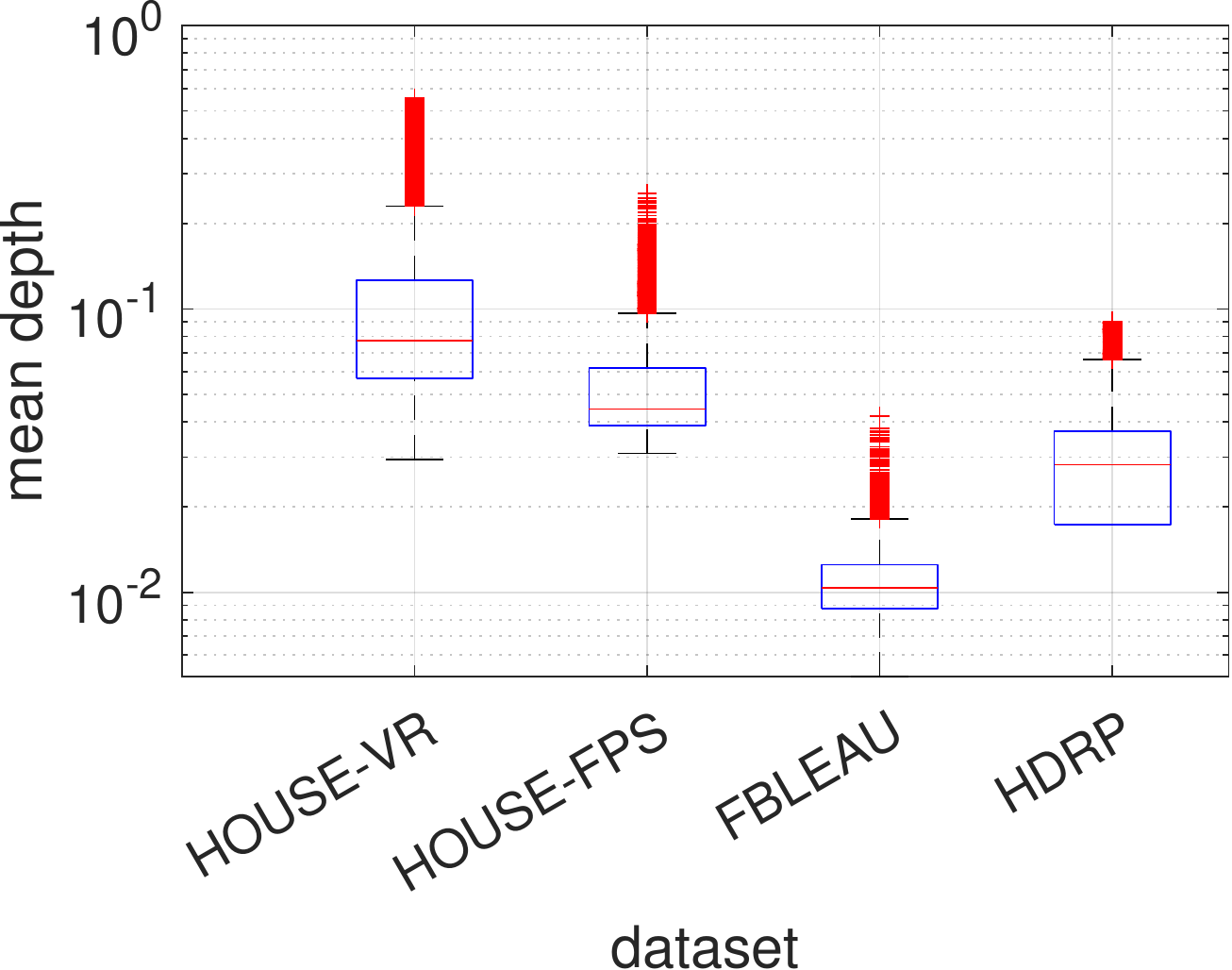}
    \label{fig:dataset_mean}
  \end{subfigure}\hspace{5mm}
  \begin{subfigure}[b]{0.4\linewidth}
   \includegraphics[width=1\linewidth]{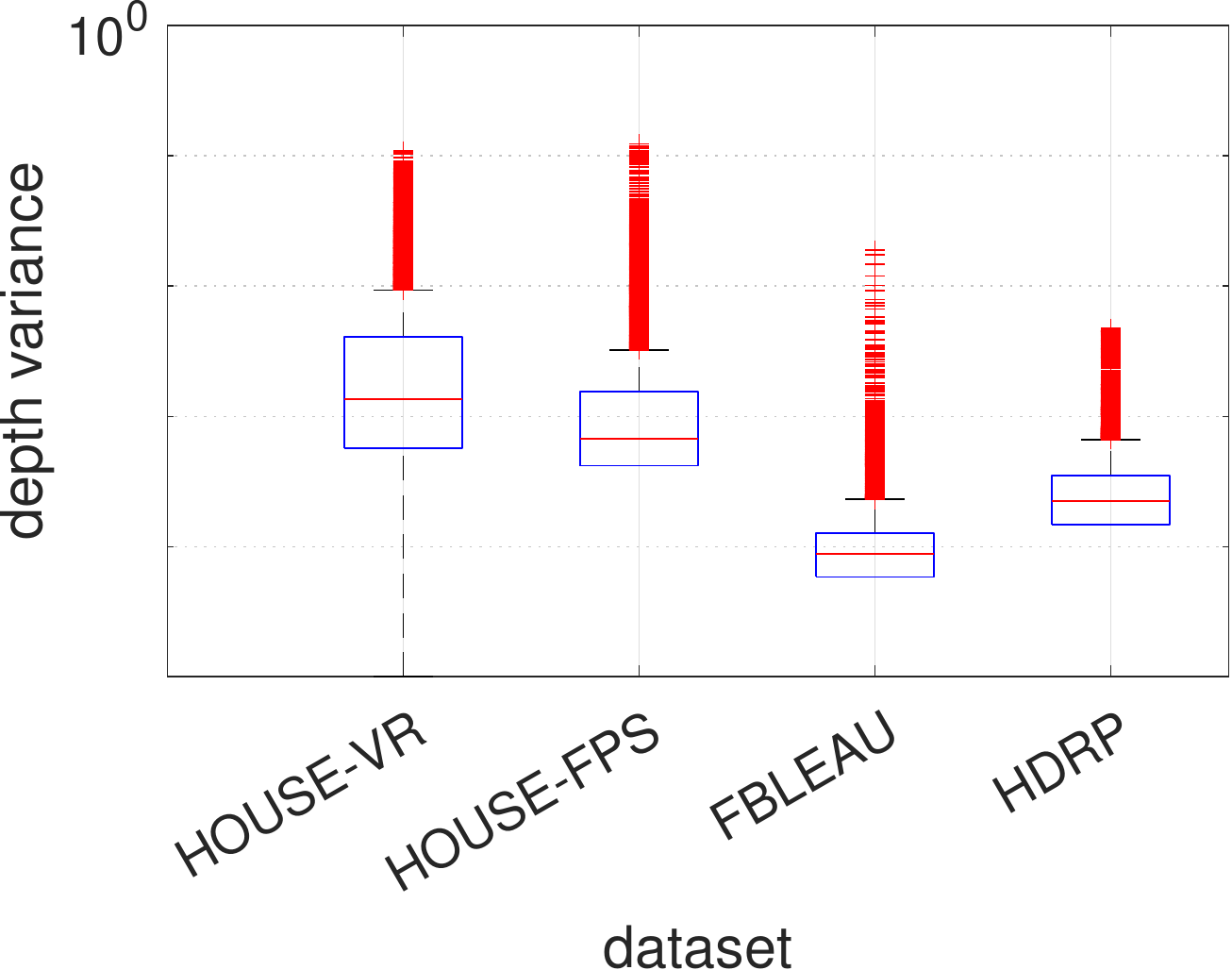}
    \label{fig:dataset_var} 
    \end{subfigure}
  \caption{Dataset statistics.}
  \label{fig:dataset_stats}  
\end{figure}

We used the datasets described in Section \ref{subsec:datasets} to train the model. Figure \ref{fig:dataset_stats} shows distribution of the mean and variance of depth per depth map in the four datasets. There are substantial differences, particularly the FBLEAU dataset differs stands out with low mean depth values which refer to distant objects as the depth buffer values are inverted. Indeed, it is an outdoor scene, whereas the others are indoor. Datasets were split into non-overlapping train (80\%) and test (20\%) data. When multiple datasets were used to train a model, we included an equal number of samples from each one.

We noticed that smaller ResNet configurations, such as ResNet-11, could be properly trained from scratch using our depth map datasets, but the models overfit with the ResNet-50 (or larger). Therefore, we instead use ResNet-50 pretrained on the ImageNet dataset (i.e., RGB images) and fine tune it with our depth map datasets to generate all the models presented in the next section. 

We used the RAdam optimizer with default values ($lr$=0.001, $betas$=(0.9, 0.999), $eps$=1e-8). Learning rate was decayed by a factor of 0.1 every 7 epochs, batch size set to 10, and each model trained for 30 epochs.

\subsection{Results}


\begin{figure}[ht] 
\centering
  \begin{subfigure}[b]{0.32\linewidth}
     \includegraphics[width=1\linewidth]{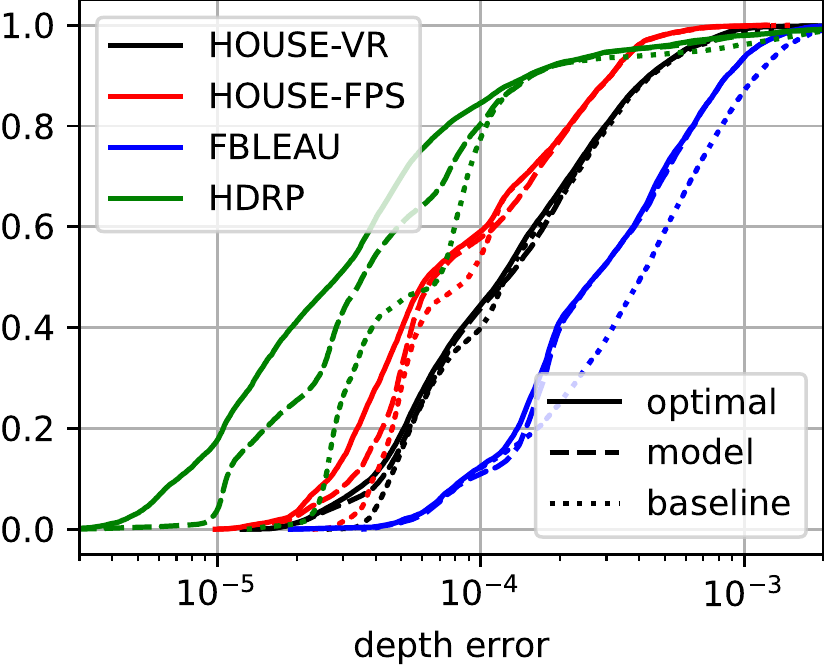}
    \caption{Two-game model (2GM)}
    \label{fig:model_error_twoscenes}
  \end{subfigure}
  \begin{subfigure}[b]{0.32\linewidth}
   \includegraphics[width=1\linewidth]{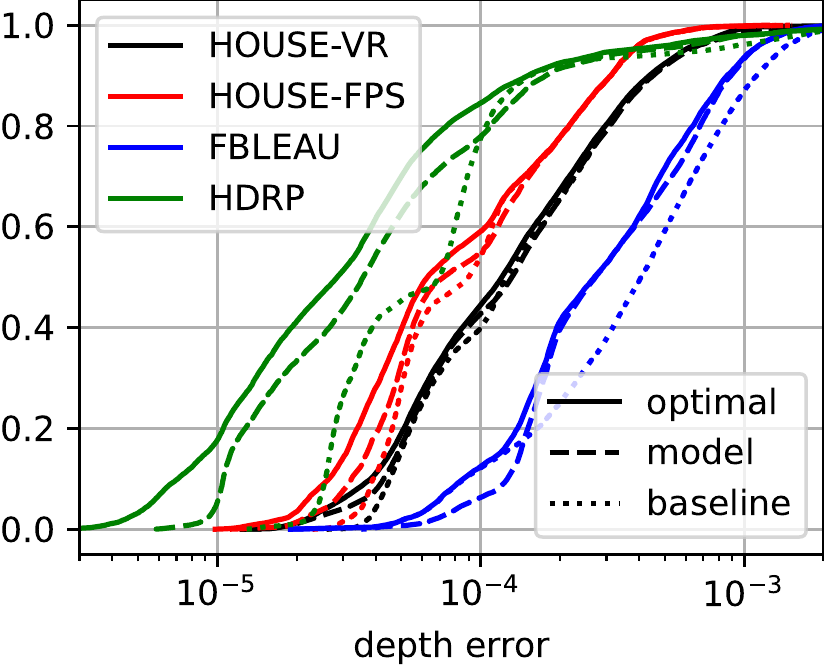}
    \caption{Three-game model (3GM)} 
    \label{fig:model_error_threescenes} 
    \end{subfigure}
  \begin{subfigure}[b]{0.32\linewidth}
   \includegraphics[width=1\linewidth]{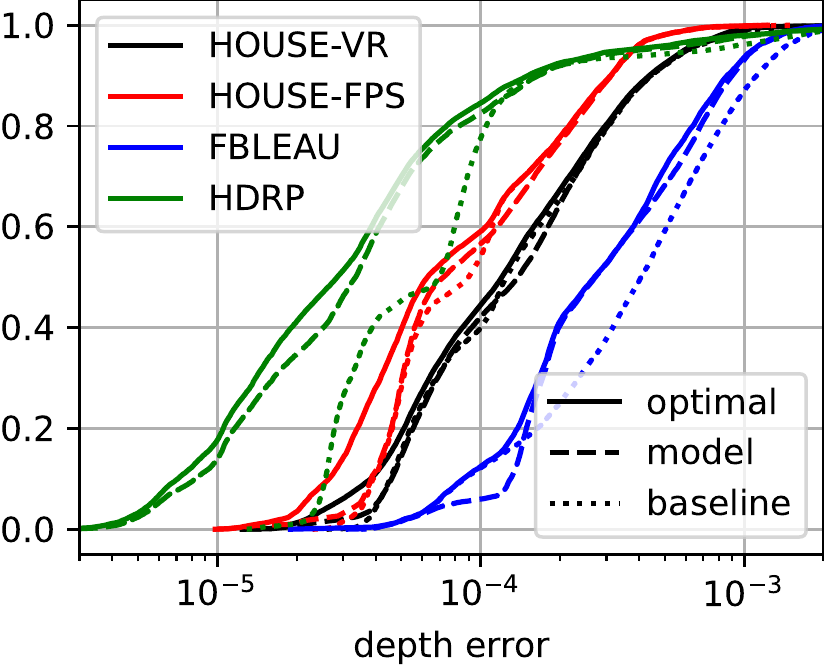}
    \caption{3GM trained on 256x256 input} 
    \label{fig:model_error_256} 
    \end{subfigure}
  \caption{Distribution of depth error.}
  \label{fig:model_error}  
\end{figure}

\begin{figure}[ht] 
\centering
  \begin{subfigure}[b]{0.32\linewidth}
     \includegraphics[width=1\linewidth]{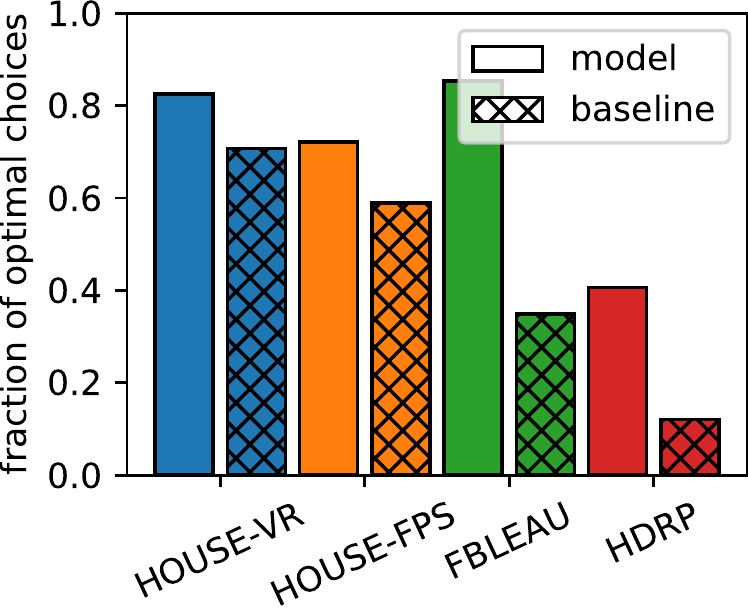}
    \caption{Two-game model}
    \label{fig:nb_optimal_choices_twoscene}
  \end{subfigure}
  \begin{subfigure}[b]{0.32\linewidth}
   \includegraphics[width=1\linewidth]{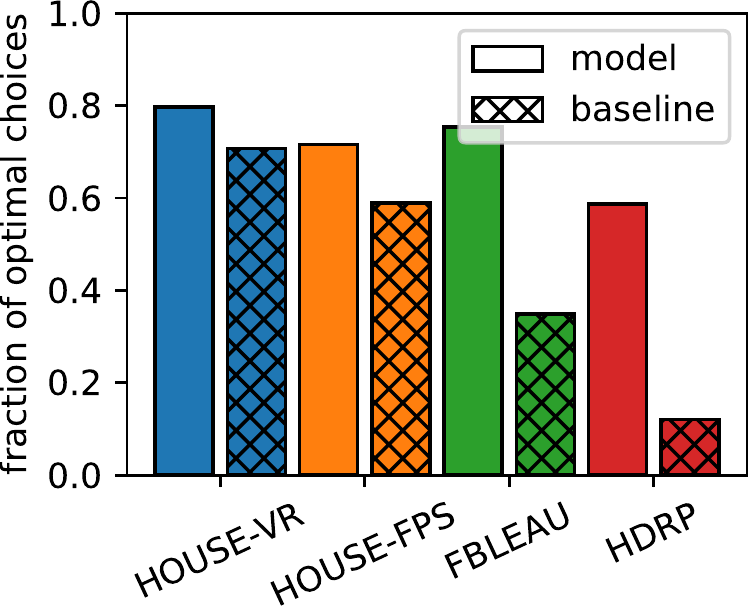}
    \caption{Three-game model} 
    \label{fig:nb_optimal_choices_threescene} 
    \end{subfigure}
  \begin{subfigure}[b]{0.32\linewidth}
   \includegraphics[width=1\linewidth]{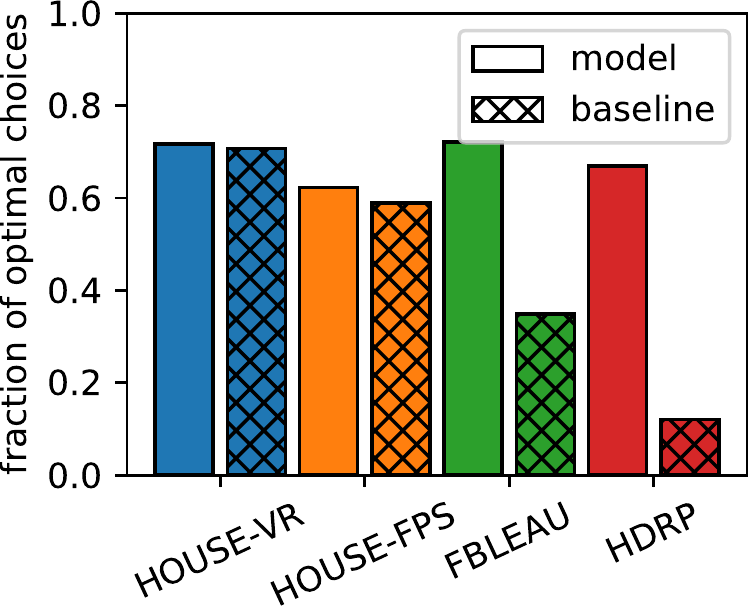}
    \caption{3GM trained on 256x256 input} 
    \label{fig:nb_optimal_choices_256} 
    \end{subfigure}
  \caption{Number of optimal choices.}
  \label{fig:nb_optimal_choices}  
\end{figure}


In presenting the results, we mainly focus on the resulting depth error and deliberately avoid using an application specific metric, such as an objective image quality metric applied to warped depth images.
Figure \ref{fig:model_error_twoscenes} shows depth encoding error results with a model fine-tuned on a combination of HOUSE-VR and FBLEAU datasets. We call this model the \textit{two-game model} as it contains data from two different games. The error was calculated by choosing for each depth map the precision that yields the lowest error according to the model. The baseline error was calculated by using the best performing precalculated configuration for each bitrate with the HOUSE-VR dataset, i.e., 12-bit precision for bitrates less than 50Mbps and 14-bit precision otherwise (Fig. \ref{fig:mean_errors_vbp_444}). Figure \ref{fig:nb_optimal_choices_twoscene} shows the fraction of optimal choices by the model and baseline on the different datasets.

The model yields nearly optimal decisions for the datasets that were used in training. The baseline performs almost as well with the HOUSE-VR that it was extracted from. However, we notice clearer differences with the other datasets. In particular, the model beats the baseline on the HOUSE-FPS dataset which suggests that it generalizes across different trajectories of the same dataset, as HOUSE-FPS was not included in training. Moreover, compared to the baseline, the model shows much better, although not quite optimal, accuracy on the HDRP dataset, representing a game scene not included in the training. 


We also trained a three-game model using a combination of all the three different game scenes, i.e. HOUSE-VR, FBLEAU, and HDRP datasets. The accuracy of the model clearly increases on the HDRP test data compared to the the two-game model, while there is minor penalty with the test data from the other datasets (Figures~\ref{fig:model_error_threescenes} and \ref{fig:nb_optimal_choices_threescene}).

\begin{figure}[th] 
 \centering
 \includegraphics[width=0.7\columnwidth]{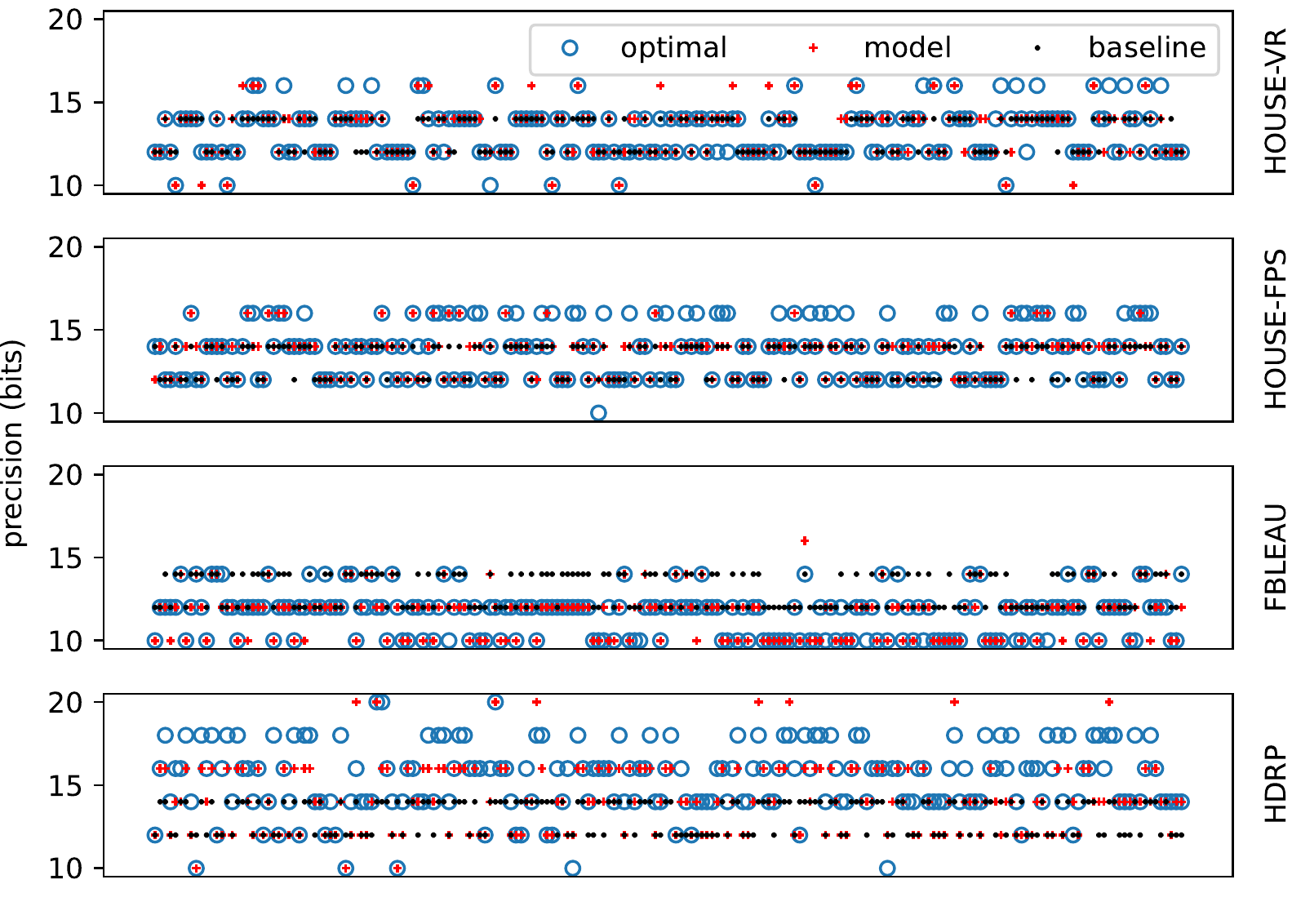}
 \caption{Chosen precision for first 200 depth maps with the two-game model.}
 \label{fig:choices}
\end{figure}

Figure \ref{fig:choices} visualizes the precision chosen by the model and baseline compared to the optimal for the first 200 depth maps. There are clear differences between the datasets in terms of the optimal precision and the way it varies. The model is able to adapt reasonably well, whereas a fixed baseline policy obviously is not.


We also trained a model using lower resolution depth maps as input. In this way, the model learns to predict higher resolution (512x512) depth map encoding error using smaller resolution (256x256) input, which is advantageous from inference speed perspective (Section~\ref{sec:benchmarks}). Figures~\ref{fig:model_error_256} and \ref{fig:nb_optimal_choices_256} shows that the results are mostly similar to those with 512x512 input with some exceptions, such as the overall error surprisingly being smaller on the HDRP dataset.

In case different encoded depth map resolutions are required, the model could possibly be extended for multi-resolution inference by adding resolution as an additional input to the fully connected layer at the end of the network and training it on multiple resolution datasets. Alternatively, one can train separate models for each resolution needed. Otherwise, the evaluation results suggest that the model can cope with different game scenes and camera trajectories, provided that the training data is sufficiently diverse. This is important for deployment considerations as it reduces the need to fine tune the model for each different game and/or scene.

%% file: 06_performance.tex
\section{Performance and deployment}
\label{sec:performance}
Depth maps need to be encoded in real-time in many application use cases. In remote rendering applications, the framerate requirements can range from 30-60 frames per second for cloud gaming, up to 90-120 frames per second for remote rendering of VR graphics. For this reason, the neural network which makes the decision for optimal depth map encoding settings needs to be relatively light-weight so that it does not significantly delay the complete system pipeline. In this section, we benchmark the bit depth model with different runtimes and by using reduced precision. Finally, we integrate the model into an existing game engine.

To put the results that follow into perspective, the naive alternative of repeatedly encoding a 512x512 resolution depth frame 9 times with the different precisions of the VBP scheme and then selecting the lowest depth error alternative would require roughly 10-20ms for video encoding only using hardware video encoder of a modern Nvidia graphics card. In addition, one would need to account for the decoding time as each frame would need to be decoded as many times in order to determine the depth error. This latency is infeasible for 90-120 fps real-time operation. Things would naturally become even worse with higher depthmap resolution and in the case that 1-bit precision granularity was desired in determining the optimal precision.

\subsection{Inference runtime benchmarks}
\label{sec:benchmarks}
Neural network throughput is highly dependent on the level of optimization for hardware acceleration on the used platform. Overall throughput is also dependent on the used batch size. To evaluate the inference time and throughput of our neural network model in a real-time usage scenario, we benchmarked the model with batch size 1 with a subset of the dataset (12281 depth maps), measuring the median GPU inference times for single depth map inference. Batch size of 1 mimics an application which generates depth maps on-the-fly for example as a part of a graphics pipeline.

\begin{figure}[t] 
    \centering
    \includegraphics[width=\columnwidth]{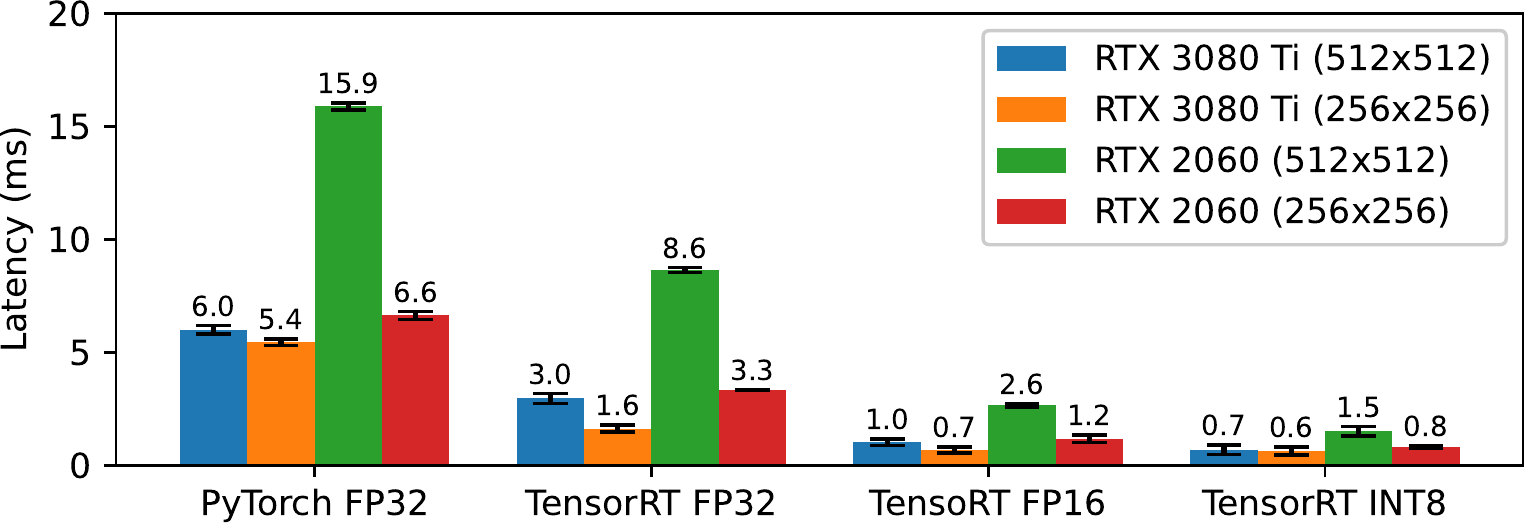}
    \caption{Median inference latencies for different runtimes, precisions and GPUs (batch size 1, depth map size 512x512 or 256x256).}
    \label{fig:latencies}
\end{figure}

The model was developed with the PyTorch framework. This framework, run on a modern desktop PC with single-precision floating-point format (FP32) is the baseline for our evaluation. We used depth map sizes of 512x512 and 256x256 as our input. The first bar group (left) in Figure~\ref{fig:latencies} shows the baseline inference latency of our model for both a currently high-end GPU (Nvidia RTX 3080 Ti) and a middle-range GPU (Nvidia RTX 2060) for the two input sizes.

The standard PyTorch runtime with full FP32 precision can run the neural network for bit depth decision in 5.4 to 6.0 ms for the RTX 3080 Ti and in 6.6 to 15.8 ms for the RTX 2060 depending on the used depth map size. This performance of the standard PyTorch runtime could be enough even for some real-time scenarios. However, as the bit depth decision model should ideally be only a small addition to the pipeline of the rest of the system, more optimized runtimes should be used possibly with reduced precision operations to accelerate the inference.

Runtimes optimized for the underlying hardware can dramatically accelerate the inference speed. As we use Nvidia GPUs, we converted our model to a TensorRT engine. TensorRT is a high-performance deep learning inference optimizer and runtime for Nvidia graphics cards~\cite{tensorrt}. Figure~\ref{fig:latencies} shows that the use of the TensorRT runtime roughly halves the inference latency and thus doubles the throughput with the same FP32 precision.

\subsection{Reduced precision}
TensorRT runtime supports quantization, where lower precision weights and activations together with lower precision math is used to speed up inference. We configured the engine to use reduced precision and re-run the benchmarks for both GPUs using half-precision floating-point format (FP16) and unified 8-bit (INT8) precision to measure the latency gains. For INT8, we also enabled FP16 in the TensoRT engine creation configuration. This allows the framework to automatically choose the best (fastest) mixed precision operation for available hardware.

The inference latency drops to 0.7/1.0 ms for FP16 using the Nvidia RTX 3080 Ti and further to 0.6/0.7 ms with INT8 precision with depth map sizes 256x256 and 512x512 respectively. Similar relative gains can be achieved using the less powerful RTX 2060 graphics card. With reduced precision both GPUs achieve inference latencies around 1 ms which is promising for applying the bit depth decision model for use in real-time applications as a part of the graphics pipeline. Reduced precision can reduce the accuracy of the model, especially with depth maps that have higher precision per channel compared to regular RGB images, but the models can be calibrated to compensate for this. In our experiments FP16 quantization did not affect the model accuracy and no calibration was needed, whereas calibration was necessary with INT8 inference. After using the built-in calibration of TensorRT, the calibrated model with INT8 inference experiences only a small loss in accuracy (1-4\% loss in the fraction of correct choices) compared to the non-quantized model. The calibration phase computes a scale value for each tensor in the network based on representative sample input values. We used 1000 input samples with batch size 60 with the \emph{MinMax Calibrator} of the TensorRT SDK.



\subsection{Game engine integration}
\label{sec:game_engine}
Depth maps are often produced one at a time for example by a graphics rendering pipeline of a real-time application. Previously, we showed that our neural network model can be run in 0.6-1.5ms on the GPU depending on the hardware and precision used. Graphics pipeline integration is however more difficult as the input needs to be fetched after the depth pass and the results of the network need to be available for the encoder as soon as possible when the frame rendering is finished. The inference needs to happen without stalling the rendering pipeline as this could lead to low framerates.

\begin{figure}[t] 
    \centering
    \includegraphics[width=\columnwidth]{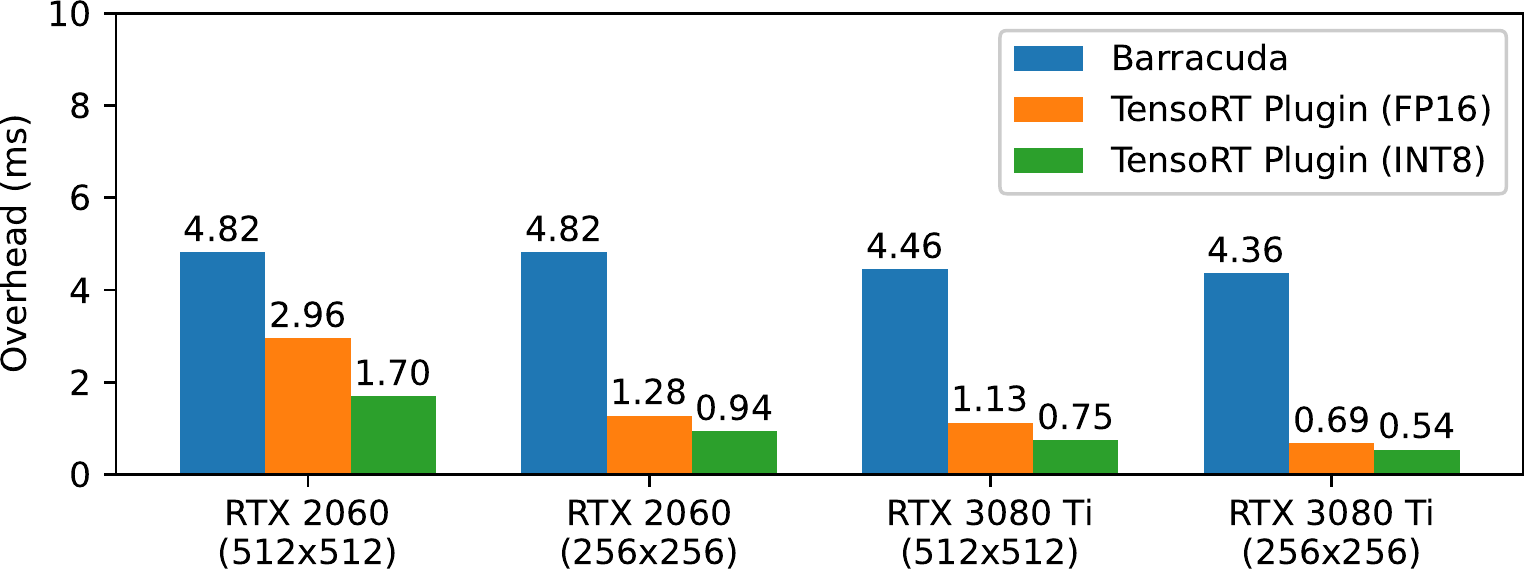}
    \caption{Overhead in frame time (lower is better) comparison between Barracuda and TensorRT plugin for neural network integration into the Unity game engine using the Nature Starter Kit 2 demo scene with different GPUs (depth map size 512x512 or 256x256).}
    \label{fig:game_engine_perf}
\end{figure}

We integrated our neural network model for bit depth decision to the popular game engine Unity~\cite{Unity}. Unity has a cross-platform Neural Networks inference library named Barracuda. It utilizes OpenGL based kernels for neural inference. We exported our model in ONNX (Open Neural Network Exchange) file format from PyTorch and imported it into Unity Barracuda. We executed the bit depth model each frame in a sample scene from a free Unity Asset \emph{Nature Starter Kit 2}~\cite{unity_nature} using 1920x1080 resolution for frame rendering and with both 512x512 and 256x256 depth frame input sizes for the neural network. In addition, we developed a native (C++) Unity plugin which uses TensorRT for inference. Figure~\ref{fig:game_engine_perf} shows the overhead per rendered frame, i.e. extra delay caused by the inference, with Unity's Barracuda implementation and with our TensorRT plugin. The Barracuda implementation has an overhead of 4.4 to 4.8 ms depending on the GPU. Our TensorRT plugin is capable of using reduced precision and has a lower overhead with only 0.5-3 ms added to the overall frame time of the application depending on the used GPU and precision. 

One possibility to further facilitate deployment with new games and game scenes would be to apply continual learning. In other words, the model would be continuously fine tuned with data that is generated while a game is being played. This could be done by, e.g., sampling every $n$ frames and adjusting $n$ so that it would not burden the GPU and video codec too much since it requires packing, encoding, and decoding the sampled depth maps multiple times.

%% file: 08_conclusion.tex
\section{Conclusion}
\label{sec:conclusion}

This paper describes our study on computer generated depth map packing for compression using a standard video codec. The use of a standard video codec was motivated by the support for hardware acceleration also on mobile devices such as phones and standalone VR/AR devices. We showed that the precision used for depth map packing has a significant impact on the resulting error given a bitrate constraint, which is caused by a combination of the packing scheme and lossy compression. Through measurements, we found that a depth map specific optimal packing configuration exists for a given bitrate. We adapt a neural network to predict the optimal depth precision and outperform a manually extracted baseline by producing near optimal predictions. Finally, we show that the used neural network model can be run in real-time on modern hardware with optimized runtimes and can be integrated into an existing game engine with very low overhead.